\documentclass[aps,floats]{revtex4}
\usepackage{amsmath,amssymb}
\usepackage{graphicx,epsfig}
\usepackage[greek,english]{babel}

\begin{document}
\bibliographystyle {plain}

\def\oppropto{\mathop{\propto}} 
\def\opsimeq{\mathop{\simeq}}
\def\opoverderline{\mathop{\overline}}
\def\operarrow{\mathop{\longrightarrow}}
\def\opsim{\mathop{\sim}}

\def\fig#1#2{\includegraphics[height=#1]{#2}}
\def\figx#1#2{\includegraphics[width=#1]{#2}}


\title{ Revisiting classical and quantum disordered systems \\
from the unifying perspective of large deviations  } 


\author{ C\'ecile Monthus }
 \affiliation{Institut de Physique Th\'{e}orique, 
Universit\'e Paris Saclay, CNRS, CEA,
91191 Gif-sur-Yvette, France}

\begin{abstract}

The theory of large deviations is already the natural language for the statistical physics of equilibrium and non-equilibrium. In the field of disordered systems, the analysis via large deviations is even more useful to describe within a unified perspective the typical events and the rare events that occur on various scales. In the present pedagogical introduction, we revisit various emblematic classical and quantum disordered systems in order to highlight the common underlying mechanisms from the point of view of large deviations.

\end{abstract}

\maketitle

\section{ Introduction  }

Just like Mr Jourdain discovering that he has been speaking in prose all his life without knowing it, physicists working in statistical physics become aware at some point that they have been using the theory of large deviations without realizing it since their very first acquaintance with the Boltzmann notion of entropy and the Gibbs theory of ensembles. 
 This language of large deviations has turned out to be very powerful
 to unify the statistical physics of equilibrium, non-equilibrium and dynamical systems
 (see the reviews \cite{oono,ellis,review_Touchette} and references therein)
and to formulate an appropriate statistical physics approach of dynamical trajectories
for various Markovian processes
(see the reviews  \cite{derrida-lecture,harris_Schu,searles,harris,mft,lazarescu_companion,lazarescu_generic}
and the PhD Theses \cite{fortelle_thesis,vivien_Thesis,chetrite_Thesis,wynants_Thesis} 
 and the HDR Thesis \cite{chetrite_HDR}).

In the field of disordered systems, the presence of random disorder variables
 induce a lot of subtle effects for the probabilities of interesting observables.
Physicists have understood from the very beginning that
some observables are non-self-averaging, i.e. their disorder-averaged value
is completely different from their typical value
(see the books \cite{luckbook,crisantibook} and references therein).
It was also realized very early that in each large typical sample, 
there will nevertheless occur rare anomalous regions of a certain size that may dominate some observables :
famous examples are the Lifshitz
essential singularities of the density of states near spectrum  edges in Anderson localization models \cite{lifshitz64,lifshitz65,lifbook,boris,luckbook}, 
the Griffiths singularities for the statics \cite{griffiths,kafri}
and the dynamics \cite{randeira,bray1,bray2} of random classical models,
 and the Griffiths phases in random quantum models (see the reviews \cite{sdrgreview,sdrgminireview} and references therein). Finally at critical points,
it was found that multifractal properties appear, for instance for the inverse participation ratios of eigenfunctions at Anderson localization transitions (see the reviews \cite{janssenrevue,mirlinrevue} and references therein) or for correlation functions in random classical spin models 
 \cite{Ludwig,Jac_Car,Cha_Ber,Cha_Ber_Sh,PCBI,BCrevue,Thi_Hil,symetriemultif}, while at Infinite Disorder fixed points,
many observables are even more broadly distributed \cite{sdrgreview,sdrgminireview}.
These few examples indicate that that the language of large deviations is even more useful in the presence of disorder
in order to describe within a unified perspective all these phenomena involving typical and rare events 
on various scales. 

The aim of the present pedagogical introduction is thus to explain to physicists
how the general theory of large deviations is the natural language to analyze the properties
of various well-known classical and quantum random models.
 It is of course not meant for mathematicians
who have been using the large deviation framework for a very long time
 (see the books \cite{ellisbook,deuschel,dembo,denHollander,saintflour,timo}
and references therein), in particular in the area of disordered systems (see the the books 
\cite{talagrand,bovier,comets}, the review \cite{zeitouni}  and references therein).
This pedagogical introduction is thus intended only for physicists who are disheartened
by the technical vocabulary used in the mathematical literature on large deviations
 (like Polish space, Borel sigma-field, cadlag function, ... ).

The following sections are organized as follows.
In section \ref{sec_1d}, we introduce the generic notations for one-dimensional random models 
and describe how observables
can be classified according to the order of the empirical property of the disorder configuration that determine them.
We then analyze the various levels of this hierarchy :
the 'Level 1' of large deviations allows to study the properties of
observables given by products of random variables (section \ref{sec_product});
the 'Level 2' of large deviations corresponds to the fluctuations of the empirical 1-point histogram
of the disorder configuration (section \ref{sec_q1}); 
the 'Level 2.5' of large deviations corresponds to the fluctuations of the empirical 2-point histogram
of the disorder configuration (section \ref{sec_q2});
finally the 'Level 3' of large deviations corresponds to the whole series of empirical histograms of arbitrary order (section \ref{sec_level3}).
In Section \ref{sec_tree}, we turn to random models defined on Cayley trees 
to analyze their properties in terms of large deviations of branches.
Our conclusions are summarized in section \ref{sec_conclusion}.
In Appendix \ref{sec_interval}, we describe an alternative classification of one-dimensional disorder configurations
 in terms of empirical intervals where the disorder remains the same.


\section{ Classification of observables in one-dimensional random models }

\label{sec_1d}


 \subsection{ Transfer-matrix formulation of one-dimensional random models}

Many classical and quantum disordered models in one dimension can be reformulated in terms of the 
product of random matrices (see the books \cite{luckbook,crisantibook} and references therein).
To have generic notations, it will be convenient to denote by $v(x)$
the disorder variable at point $x$ that is drawn independently with some probability distribution $p_{v}$ normalized to unity
\begin{eqnarray}
\sum_{v} p_v =1
\label{normapvdiscrete}
\end{eqnarray}
that should be translated into $\int dv p_v=1$ whenever the disorder $v$ is a continuous random variable.
In this paper, we have chosen to write the general equations for the case of discrete disorder $v$
 (Eq \ref{normapvdiscrete})
without the constant translation into the case of continuous disorder, but some examples of application
will involve continuous disorder.

A disorder configuration $[v(.)] \equiv [v(x )]_{x=1,2,..,L}$ on a sample of $L$ sites
occurs with the factorized probability
\begin{eqnarray}
{\mathbb P}_L  [v(.)]  \equiv 
   p_{v(1)} p_{v(2)}  ...  p_{v(L)}  = \prod_{x=1}^L  p_{v(x)}
\label{ptrajindep}
\end{eqnarray}
In this disordered sample, various physical observables can be then obtained 
by considering the product of the $L$ corresponding transfer matrices $T_{ v(x) } $
\cite{luckbook,crisantibook}. One of the most important observable is the trace of this product
\begin{eqnarray}
  {\cal T}_L [v(. )] 
\equiv  Trace \left[ T_{ v(L) }  T_{ v(L-1) } ...  T_{ v (2) } T_{ v (1) } \right]
\label{defprod}
\end{eqnarray}
The exponential growth with $L$ of its modulus $ \vert {\cal T}_L [v(. )]  \vert $
can be then measured by the finite-size Lyapunov exponent 
\begin{eqnarray}
\lambda [v(. )]   \equiv \frac{ \ln  \vert {\cal T}_L [v(. )]  \vert  }{ L} 
\label{lyapunov}
\end{eqnarray}
Of course a more complete analysis would involve the whole Lyapunov spectrum \cite{crisantibook}
of the product of matrices but will not be discussed here.

\subsection{ Statistics of the Lyapunov exponent $\lambda$ over the disorder configurations }

For large $L$, the probability distribution ${\cal P}_L(\lambda)$ of the finite-size Lyapunov exponent $\lambda$
of Eq. \ref{lyapunov}
over the disorder configurations $[v(.)]$ drawn with the probabilities of Eq. \ref{ptrajindep}
 is expected to follow the large deviation form \cite{luckbook,crisantibook} 
\begin{eqnarray}
{\cal P}_L(\lambda) \equiv \sum_{[v(.)] }  {\mathbb P}_L[v(.)]  \delta \left( \lambda - \frac{  \ln  \vert {\cal T}_L [v(. )]  \vert  }{ L}  \right) \opsimeq_{ L \to + \infty} e^{- L I(\lambda) }
\label{largedevlyapunov}
\end{eqnarray}
where $I(\lambda)$ is called the 'rate function' in the field of large deviations :
it is positive $I(\lambda) \geq 0$ and vanishes only 
at its minimum corresponding to the typical value $\lambda^{typ}$ that will be realized with probability one 
in the thermodynamic limit $L \to +\infty$.
\begin{eqnarray}
I(\lambda^{typ}) =0 = I'(\lambda^{typ})
\label{typvanish}
\end{eqnarray}
All other values $\lambda \ne \lambda^{typ}$ appear with 
a probability ${\cal P}_L(\lambda) $ that is exponentially small in $L$ in Eq. \ref{largedevlyapunov},
but they are nevertheless important to understand the behavior
of the moments of non-integer order $k$ of the trace of Eq. \ref{defprod},
as a consequence of their evaluation via the Laplace saddle-point method of the following integral over $\lambda$ 
\begin{eqnarray}
\overline{  \left\vert {\cal T}_L  [ v(. ) ] \right\vert^k }  && \equiv \sum_{[v(.)] }  P_L[v(.)]   \left\vert {\cal T}_L  [ v(. ) ] \right\vert^k
=   \int d \lambda \ {\cal P}_L(\lambda) \ e^{ L \lambda k }
\opsimeq_{ L \to + \infty} \int d \lambda \ e^{ L \left[ \lambda k  - I(\lambda) \right] } \opsimeq_{ L \to + \infty} e^{ L \phi(k) }
\label{multifz}
\end{eqnarray}
The function $\phi(k)  $ governing their exponential growth in $L$
is called the 'scaled cumulant generating function' in the field of  large deviations.
It corresponds to the Legendre transform of the rate function $I(\lambda)$ of Eq. \ref{largedevlyapunov}
as a consequence of the saddle-point evaluation of Eq. \ref{multifz}
\begin{eqnarray}
\phi(k)  && = \lambda k  - I(\lambda) 
\nonumber \\
0 && = k - I'(\lambda)
\label{legendre}
\end{eqnarray}
with the reciprocal Legendre transform
\begin{eqnarray}
I(\lambda)   && = \lambda k  - \phi(k) 
\nonumber \\
0 && = \lambda   - \phi'(k) 
\label{legendrereci}
\end{eqnarray}
For $k=0$ where $\phi(k=0) =0$ as a consequence of the normalization in Eq. \ref{multifz}, 
one obtains that the typical value $\lambda^{typ}$
where the rate function vanishes (Eq. \ref{typvanish}) corresponds to the derivative
\begin{eqnarray}
 \lambda^{typ}   = \phi'(k =0) 
\label{typq0}
\end{eqnarray}
while all moments of order $k \ne 0$ are dominated by non-typical values 
of the Lyapunov exponent in the saddle-point calculation of Eq. \ref{multifz}.

Since the typical Lyapunov exponent $\lambda^{typ}$ appear with probability one in the thermodynamical limit $L \to +\infty$, one of the main goal in the field of 
 products of random matrices has been to compute it
in various models via the Dyson-Schmidt invariant measure method \cite{luckbook,crisantibook,tourigny}.
In the present paper, our goal will be instead
 to focus on the simplest cases where the whole large deviations
rate function $I(\lambda)$ can be explicitly obtained.


\subsection{ Examples of observables corresponding to products of random variables }

\label{sec_obsprod}

It is clear that the simplest case is of Eq. \ref{defprod} is 
when the transfer matrices $T_v $ are replaced by numbers $t_v$
\begin{eqnarray}
  \tau_L [v(. )] =  t_{ v(L) }  t_{ v(L-1) } ...  t_{ v (2) } t_{ v (1) } =\prod_{x=1}^L  t_{v(x)}
\label{prodrandom}
\end{eqnarray}
This case occurs in various disordered models, either exactly or approximately in some region of parameters,
as displayed by the following examples.

\subsubsection{ Examples of observables that are exactly given by products of random variables }

(1-a) In the classical Ising chain with random couplings $J(x)$,
the two-spin correlation function reads \cite{Der_Hil_corre,luckbook,crisantibook}
 \begin{eqnarray}
C(x_0,x_0+r) = \prod_{x=x_0 }^{x_0+r-1} \tanh (\beta J(x) )
\label{cr}
\end{eqnarray}

(1-b) In the random quantum spin chains corresponding to free majorana fermions, 
the possible edge Majorana zero modes that characterize the topological phases
are given in terms of product of random variables in the simplest cases (see \cite{c_maj}
and references therein for various examples).


\subsubsection{ Observables that can be approximated by products of random variables in certain regions of parameters }

(2-a) For the Anderson Localization tight-binding model with hopping $V$ and random on-site-energy $\epsilon(x)$,
the eigenfunction $\psi_{x_0} $ localized on site $x_0$ for $V=0$ can be approximated at lowest order in the hopping $V$ 
in the so-called Forward Approximation \cite{alt_levitov,luca,forward,c_mblfss} by the product
\begin{eqnarray}
 \psi_{x_0} (x_0+r) && \simeq  \prod_{x=x_0+1 }^{x_0+r} \left(\frac{V}{\epsilon(x_0)-\epsilon(x) } \right)
\label{psi0r}
\end{eqnarray}

(2-b) For the quantum Ising chain with random couplings $J(x)$ and random transverse fields $h(x)$,
the two-spin correlation function is given at lowest order in perturbation in the couplings by the product
 \begin{eqnarray}
C(x_0,x_0+r) = J(x_0) \prod_{x=x_0+1 }^{x_0+r-1} \frac{J(x)}{h(x)}
\label{crq}
\end{eqnarray}
This form can also be understood from the Strong Disorder RG approach \cite{sdrgreview,sdrgminireview} when only sites are decimated, or from the Cavity approach \cite{cavity1,cavity2,cavity3}.


\subsection{ Classification of observables in terms of empirical histograms of the disorder configuration }

For each disorder configuration $[v(x)]_{x=1,2,..,L}$ with periodic boundary conditions $v(L+x) = v(x )$,
the empirical 1-point histogram 
\begin{eqnarray}
Q_{v_1 }  [ v (.) ] \equiv \frac{1}{L}  \sum_{x=1}^{L} \delta_{ v_1, v (x+1 )} 
\label{q1}
\end{eqnarray}
measures the frequencies of the possible values $v_1$ of the disorder variable.
More generally, the empirical r-point histogram 
\begin{eqnarray}
Q_{v_r ... v_2 v_1}  \left[ v (.) \right]  \equiv \frac{1}{L}  \sum_{x=1}^{L} 
\delta_{ v_r, v (x+r )}  ...
\delta_{ v_2, v (x+2 )} \delta_{ v_1, v (x+1 )}
\label{qr}
\end{eqnarray}
measures the frequencies of the occurrence of the r consecutive values $(v_r,...v_2,v_1)$ in the disordered sample.
This hierarchy can be constructed up to the maximal value $r_{max}=L$ that corresponds to the total length $L$ of the disorder configuration
\begin{eqnarray}
Q_{v_L v_{L-1} ... v_2 v_1}  \left[ v (.) \right]  \equiv \frac{1}{L}  \sum_{x=1}^{L} 
\delta_{ v_L, v (x+L )} \delta_{ v_{L-1}, v (x+L-1 )} ...
\delta_{ v_2, v (x+2 )} \delta_{ v_1, v (x+1 )}
\label{qrmax}
\end{eqnarray}
i.e. this represents the average over the $L$ translations via $x=1,2,..,L$ of the
initial disorder configuration.

The observables of the disordered models can be then classified according to the order $r$ of the empirical r-point histogram that allows to reconstruct them.
For instance, the product of Eq. \ref{prodrandom} can be rewritten in terms of the empirical 1-point histogram 
$Q_{v_1} [v (.) ] $ of Eq. \ref{q1} as
\begin{eqnarray}
  \tau_L [v(. )] = \prod_{x=1}^L  t_{v(x)} = \prod_{v_1} \left( t_{v_1} \right)^{L Q_{v_1} [v (.) ] }
\label{prodrandomq1}
\end{eqnarray}
The physical interpretation is that the product of random variables
is not sensitive to the order of appearance of the disorder variables $v(x)$, but depends only on the 
global frequencies of the possible values $v_1$ that are summarized in the empirical 1-point histogram 
$Q_{v_1} [v (.) ] $.

An example of observable that depends only on the empirical r-point histogram of Eq. \ref{qr}
is the Spatial-Average within a given sample 
of the 2-point correlation function at distance $r$ of Eq. \ref{cr}
 in a given sample 
\begin{eqnarray}
C^{SpAv}(r) \equiv \frac{1}{L}  \sum_{x=1}^{L} C(x,x+r) 
=
\frac{1}{L}  \sum_{x=1}^{L} \prod_{y=x }^{x+r-1} \tanh (\beta J(y) )
= \sum_{J_r} ... \sum_{J_1} \left( \prod_{j=1 }^{r} \tanh (\beta J_j )  \right) Q_{J_r ... J_2 J_1}  \left[ J (.) \right] 
\label{cravtrans}
\end{eqnarray}
whose statistics is discussed in \cite{Der_Hil_corre} 
to stress that it will coincide with the disorder-averaged correlation function only for the small sizes $r \leq (cst) \ln L $.
Finally, the most general observables depend on the empirical L-point histogram of Eq. \ref{qrmax}
that contains the complete information on the disorder configuration.

The usefulness of this classification is that once one has identified that an observable $ A[v(.)]$
depends on the disorder configuration $ [v(.)]$ only via its empirical r-point histogram $Q_{(rpoints)}  \left[ v (.) \right]$
of Eq. \ref{qr}
\begin{eqnarray}
A[v(.)] = {\cal A} \left( Q_{(rpoints)}  \left[ v (.) \right] \right)
\label{atoaempi}
\end{eqnarray}
 then its probability distribution over the disorder configurations drawn with Eq. \ref{ptrajindep}
depends only on the probability distribution $P_L [ Q_{(rpoints)}  ]  $ of the empirical r-point histogram
\begin{eqnarray}
{\cal P}_L(A) \equiv 
\sum_{[v(.)] }  {\mathbb P}_L[v(.)]  \delta \left( A -  {\cal A} \left( Q_{v_r ... v_2 v_1}  \left[ v (.) \right] \right)  \right) 
= \sum_{[Q_{(rpoints)}] } P_L [ Q_{(rpoints)}  ]   \delta \left( A -  {\cal A} \left( Q_{(rpoints)}   \right)  \right) 
\label{pdfaqr}
\end{eqnarray}
In the theory of large deviations, it turned that
the probability distributions $P_L [ Q_{(rpoints)}  ]  $
of the empirical r-point histograms of various order $r$ have been labelled by levels 
as follows \cite{oono,review_Touchette} :
the Level 2 corresponds to the empirical 1-point histogram $Q_.$,
the Level 2.5 corresponds to the empirical 2-point histogram $Q_{..}$,
the Level 3 corresponds to the full hierarchy of arbitrary $r$ up to the limit $r \to +\infty$.
In the following sections, we will thus describe this hierarchy,
starting with the Level 1 that corresponds to the large deviations properties of sums of random variables,
that are important to fully characterize the statistics of products of random variables.


\section{ Product of random variables
as the level-1 of large deviations }

\label{sec_product}

In this section, we focus on the product of random variables corresponding to the modulus of 
 Eq \ref{prodrandom}
\begin{eqnarray}
 \vert \tau_L [v(. )] \vert=  \prod_{x=1}^L  \vert t_{v(x)} \vert 
\label{prodrandompositive}
\end{eqnarray}
and on the corresponding finite-size Lyapunov exponent of Eq. \ref{lyapunov}
\begin{eqnarray}
\lambda [v(. )]   \equiv \frac{ \ln  \vert \tau_L [v(. )]  \vert  }{ L} = \frac{1}{L} \sum_{x=1}^L  \ln \vert t_{v(x)} \vert
\label{lyapunovprod}
\end{eqnarray}
As explained in detail in the previous section, this is the simplest problem that occur in the field of disordered systems.
In the language of large deviations, the properties of the sum of random variables of Eq. \ref{lyapunovprod}
is also the simplest example corresponding to the so-called 'Level-1' description \cite{oono,review_Touchette}.

\subsection{ Moments of non-integer order $k$  }

The moments  of non-integer order $k$ of the product in Eq. \ref{prodrandompositive}
can be directly computed as a consequence of the independence of the disorder variables $v(x)$ on the $L$ sites 
(Eq. \ref{ptrajindep})
\begin{eqnarray}
\overline{  \left\vert \tau_L  [ v(. ) ] \right\vert^k }  && 
= \overline{ \prod_{x=1}^L  \vert t_{v(x)} \vert^k } = \left[ \overline{ \vert t_{v} \vert^k }   \right]^L
\label{momentsk}
\end{eqnarray}
So the scaled cumulant generating function $\phi(k)  $ governing their exponential growth in $L$ (Eq \ref{multifz})
is given, actually even for any finite $L$,  by the simple expression
\begin{eqnarray}
\phi(k) = \frac{  \ln \left( \overline{  \left\vert \tau_L  [ v(. ) ] \right\vert^k  } \right) } {L} 
= \ln \left[ \overline{ \vert t_{v} \vert^k }   \right]
\label{phikp}
\end{eqnarray}
in terms of the moments $\overline{ \vert t_{v} \vert^k }  $ of the elementary variable $\vert t_{v} \vert $.

\subsection{ Rate function $I(\lambda)$ governing the large deviations of the Lyapunov exponent $\lambda$  }

The rate function $I(\lambda)$ governing the large deviations (Eq \ref{largedevlyapunov})
of the Lyapunov exponent $\lambda$ of Eq. \ref{lyapunovprod}
can be computed either directly if the probability distribution of the sum of Eq. \ref{lyapunovprod} is known
or it can be obtained via the reciprocal Legendre transform (Eq. \ref{legendrereci}) from the knowledge of the function $\phi(k)$ of Eq. \ref{phikp}.
Let us now recall some simple examples that will be useful later (in section \ref{sec_tree}).

\subsection{ Examples for the equilibrium of disordered classical models  }

\label{subsec_eqbeta1d}

In the field of disordered classical models, the simplest example is when the variable $t_{v(x)}$
corresponds to the Boltzmann weight at inverse temperature $\beta$ of the random potential $v(x)$
\begin{eqnarray}
  t_{v(x)} = e^{\beta v(x) }
\label{boltz1}
\end{eqnarray}
Then Eq. \ref{prodrandompositive} represents the Boltzmann weight of the $L$ sites
\begin{eqnarray}
  \tau_L [v(. )] =  \prod_{x=1}^L   e^{\beta v(x) }
\label{boltzl}
\end{eqnarray}
and Eq. \ref{lyapunovprod} corresponds to the energy per site (up to the factor $\beta$)
\begin{eqnarray}
\lambda [v(. )]   = \beta \frac{1}{L} \sum_{x=1}^L  v(x)
\label{lyapunovbolt}
\end{eqnarray}

For instance if the distribution of the potential $v$ is Gaussian of zero mean
\begin{eqnarray}
p^{Gauss}(v) = \frac{1}{ \sqrt{ 2 \pi \sigma^2} } e^{ - \frac{v^2}{2  \sigma^2} }
\label{gauss}
\end{eqnarray}
then both the rate function $I(\lambda)$ and the scaled cumulant generating function $\phi(k)  $ are simply quadratic
\begin{eqnarray}
I^{Gauss}(\lambda) && =  \frac{\lambda^2}{2 \beta^2  \sigma^2} 
\nonumber \\
\phi^{Gauss}(k) && =  \frac{ k^2 \beta^2 \sigma^2}{2} 
\label{largedevgauss}
\end{eqnarray}
Another example is when the distribution of the potential $v$ is the Bernoulli distribution
\begin{eqnarray}
p^{Bernoulli}_v = p \delta(v-v_0) + (1-p) \delta(v)
\label{bernouilli}
\end{eqnarray}
then the rate function $I(\lambda)$ and the scaled cumulant generating function $\phi(k)  $ read
\begin{eqnarray}
I^{Bernoulli}(\lambda) && = \frac{\lambda}{\beta v_0} \ln \left( \frac{\lambda}{p \beta v_0}\right) 
+ \left( 1- \frac{\lambda}{\beta v_0} \right) \ln \left( \frac{ 1- \frac{\lambda}{\beta v_0} }{1-p} \right)
\nonumber \\
\phi^{Bernoulli}(k) && =  \ln \left[ p e^{ k \beta  v_0  } + (1-p)    \right]
\label{largedevbernouilli}
\end{eqnarray}
So it is important to stress here that
 the large deviations properties depend on all the details of the disorder distribution $p_v$,
in contrast to the small deviations region described by the Central-Limit-Theorem 
that corresponds to the expansion at lowest order of the rate function $I(\lambda)$ around its vanishing minimum
at the typical value $\lambda_{typ}$ of Eq. \ref{typvanish}
\begin{eqnarray}
I^{CLT}(\lambda) && = \frac{I''(\lambda_{typ}) }{2} ( \lambda-\lambda_{typ})^2 + o(( \lambda-\lambda_{typ})^2 ) 
\label{clt}
\end{eqnarray}

\subsection{ Examples for disordered quantum models  }

\label{exampleAnderson}

For the Anderson Localization model in the Forward approximation of Eq. \ref{psi0r},
it is usual to consider the box distribution of width $(2W)$ for the random on-site energy $\epsilon(x)$
\begin{eqnarray}
p^{Box} (\epsilon) = \frac{ \theta( -W \leq \epsilon \leq W )}{2 W}
\label{box}
\end{eqnarray}
The elementary variable $ t_{\epsilon(x)} $ in the product in Eq. \ref{psi0r} at the center of the band $ \epsilon(x_0)=0$
\begin{eqnarray}
  t_{\epsilon(x)} = \frac{V}{ \vert \epsilon(x) \vert} 
\label{tanderson}
\end{eqnarray}
has then moments only in the region $k<1$
\begin{eqnarray}
 \overline{ \vert t_{\epsilon} \vert^k }  = V^k \int_0^{W} \frac{d \epsilon}{W} \epsilon^{-k} 
= \left( \frac{ V}{ W}\right)^k \frac{1}{1-k} 
\label{momentskbox}
\end{eqnarray}
So the scaled cumulant generating function $\phi(k)  $ of Eq. \ref{phikp} reads
\begin{eqnarray}
\phi^{Anderson}(k) = \ln \left[ \overline{ \vert t_{\epsilon} \vert^k }   \right] = k \ln \left( \frac{ V}{ W}\right) - \ln (1-k)
\label{phikanderson}
\end{eqnarray}
with the corresponding rate function
\begin{eqnarray}
I^{Anderson}(\lambda)= \lambda - \ln \left( \frac{ V}{ W}\right) -1 -\ln \left[ \lambda - \ln \left( \frac{ V}{ W}\right) \right]
\label{ianderson}
\end{eqnarray}


\section{ Empirical 1-point histogram as the level-2 of large deviations  }

\label{sec_q1}

In this section, we focus on the probability of the empirical 1-point histogram of Eq. \ref{q1}
over the disorder configurations $v(.)$ drawn with Eq. \ref{ptrajindep}
\begin{eqnarray}
P_L [ Q_{.}  ]    \equiv \sum_{[v(.)] }  {\mathbb P}_L[v(.)]  \prod_{v_1} \delta \left(Q_{v_1 } - \frac{1}{L}  \sum_{x=1}^{L} \delta_{ v_1, v (x+1 )}   \right)
\label{defprobaq1}
\end{eqnarray}
Of course the typical value of this histogram is the 'true' probability distribution $p_v$ of the disorder
 (Eq. \ref{normapvdiscrete})
\begin{eqnarray}
Q_v^{typ} = p_v
\label{q1typ}
\end{eqnarray}
but here the goal is to describe its fluctuations for large $L$.
In the language of large deviations \cite{oono,ellis,review_Touchette}, this is known as
 the 'Level-2 description of the empirical measure'.
The essential result is the large deviation form for large $L$
\begin{eqnarray}
P_L [ Q_{.}  ]   \opsimeq_{L \to +\infty} C_1 [ Q_. ]
e^{-L S^{rel} (Q_{.}  \vert p_.  )  }
\label{sanov}
\end{eqnarray}
where 
\begin{eqnarray}
C_1 [ Q_. ] = \delta \left( 1-  \sum_{v}  Q_{v}    \right)
\label{c1}
\end{eqnarray}
represents the normalization constraint of the empirical histogram 
(the notation $\delta(Y)$ represents the discrete Kronecker symbol $\delta_{0,Y}$
but has been chosen here for better readability of the argument $Y$),
while the rate function is the relative entropy of the empirical 1-point histogram  $Q_{v}$
with respect to the true probability distribution $p_{v}$ of the disorder
\begin{eqnarray}
 S^{rel} (Q_{.}  \vert p_.  )  \equiv \sum_{v} Q_{v}   \ln \frac{Q_{v}  }{p_{v}   }
\label{s1relative}
\end{eqnarray}
This result is known as the Sanov theorem in the field of large deviations \cite{oono,ellis,review_Touchette}
and can be considered as the true cornerstone of the whole theory,
with many further generalizations for the higher levels.
It is thus important to fully understand its origin and its physical meaning,
via the three following different derivations.

\subsection{ First approach via the multinomial distribution  }

Since each disorder value $v(x) $ is drawn with probability $p_{v(x)}$ independently on each of the $L$ sites $x=1,2,..,L$ (Eq \ref{ptrajindep}), the probability of the empirical 1-point histogram $Q_{.} $
of Eq. \ref{q1} amounts to analyze the integer numbers $(L Q_{v})$ of the occurrences of each value $v$ 
and is thus given by the multinomial distribution
\begin{eqnarray}
P_L [ Q_{.}  ]   = \delta \left( 1-  \sum_{v} Q_{v}    \right) \frac{ L!}{ \displaystyle \prod_{v} (L Q_{v})!   }  \prod_{v}
 [ p_{v} ]^{L Q_{v}}
\label{multinomial}
\end{eqnarray}

The Stirling's approximation for the factorials $m! \simeq \sqrt{2 \pi m} \ m^m e^{-m}$
then yields the large deviation form of Eq. \ref{sanov} with the relative entropy of Eq. \ref{s1relative}.
This derivation based on the application of the Stirling's approximation to the multinomial distribution of Eq. \ref{multinomial} goes back to Boltzmann \cite{ellis} and appears in all statistical physics lectures.

\subsection{ Second approach via the generating function }

\label{sec_genelevel2}

Another derivation is based on the generating function of the empirical 1-point histogram of Eq. \ref{defprobaq1}
\begin{eqnarray}
 G_L [ \nu_{.}  ]  && 
\equiv  \sum_{ Q_.} P_L [ Q_{.}  ]  \ e^{ \displaystyle  L \sum_{v} \nu_{v} Q_{v} }   
 =  \sum_{ v(1)}... \sum_{v(L)} 
p_{v(1)} p_{v(2)}  ...  p_{v(L)}
 \  e^{ \displaystyle   \sum_{v} \nu_{v}  \sum_{x=1}^L \delta_{v,v(x)}  }   
 \nonumber \\ &&
 = \prod_{x=1}^L \left(  \sum_{v(x)} p_{v(x)} e^{ \nu_{v(x)}  }   \right)
= \left(  \sum_{v} p_{v} e^{ \nu_{v}  }     \right)^L
\label{geneq1}
\end{eqnarray}
This factorized form is valid already for any finite $L$
and the corresponding scaled cumulant generating function $\Phi [ \nu_{.}  ]$
governing the exponential growth with $L$
\begin{eqnarray}
 G_L [ \nu_{.}  ]   && =  e^{ L \Phi [ \nu_{.}  ]  }
\label{geneq1phi}
\end{eqnarray}
is given in terms of the generating function of the disorder distribution $p_{v}$
\begin{eqnarray}
 \Phi [ \nu_{.}  ]  && = \ln \left(  \sum_{v} p_{v} e^{ \nu_{v}  }   \right)
\label{phi1}
\end{eqnarray}
where the analogy with Eq. \ref{phikp} is clear.
It is now useful to show the link with the the relative entropy of Eq. \ref{s1relative}
via the Legendre transform and the reciprocal Legendre transform respecify.

\subsubsection{ Link with the relative entropy via the Legendre transform }

The generating function of Eq \ref{geneq1} 
can be rewritten in terms of Eq. \ref{sanov}  as
\begin{eqnarray}
 G_L [ \nu_{.}  ]  && 
\equiv  \sum_{ Q_.} P_L [ Q_{.}  ]  e^{ \displaystyle  L \sum_{v} \nu_{v} Q_{v} }  
\opsimeq_{L \to +\infty} 
 \sum_{ Q_.}\delta \left( 1-  \sum_{v} Q_{v}    \right) 
e^{ \displaystyle  L \left[ \sum_{v} \nu_{v} Q_{v} 
- S^{rel} (Q_{.}  \vert p_.  )     \right] }  
\label{geneq1saddle}
\end{eqnarray}
The Laplace's saddle point method for large $L$ yields that one should optimize over $Q_{.}$ the 
function in the exponential
 in the presence of the normalization constraint 
$\left( 1-  \sum_{v} Q_{v}    \right)  $ in order to obtain the function $\Phi [ \nu_{.}  ]  $ of Eq. \ref{geneq1phi}
\begin{eqnarray}
 \Phi [ \nu_{.}  ]  && 
=  \max \limits_{ Q_{.} : 1= \sum_{v} Q_{v}  } \left[ \sum_{v} \nu_{v} Q_{v} 
- S^{rel} (Q_{.}  \vert p_.  )   \right]
=  \max \limits_{ Q_{.} : 1= \sum_{v} Q_{v}  } \left[ \sum_{v} \nu_{v} Q_{v} 
- \sum_{v} Q_{v}   \ln \frac{Q_{v}  }{p_{v}   }
  \right]
\label{phi1srel}
\end{eqnarray}
Taking into account the constraint via some Lagrange multiplier $\eta$, 
one needs to optimize the functional
\begin{eqnarray}
 {\cal L}(Q_.) = \sum_{v} Q_{v} 
\left[ \nu_{v}-    \ln \frac{Q_{v}  }{p_{v}   } \right] +\eta  \left( 1-  \sum_{v} Q_{v}    \right)
\label{funct1}
\end{eqnarray}
over the values $Q_{v}$
\begin{eqnarray}
0= \frac{ \partial   {\cal L}(Q_.) }{\partial Q_{v} } = 
 \nu_{v}-    \ln \frac{Q_{v}  }{p_{v}   } -1 - \eta  
\label{funct1deri}
\end{eqnarray}
One obtains the optimal solution
\begin{eqnarray}
Q^*_{v} = p_{v} e^{\nu_{v} -1 - \eta  }
\label{q1star}
\end{eqnarray}
where the Lagrange multiplier $\eta$ is fixed by the constraint
\begin{eqnarray}
 1 =  \sum_{v} Q^*_{v}  = e^{ -1 - \eta} \sum_{v} p_{v} e^{\nu_{v}   }
\label{lagrangemu1}
\end{eqnarray}
The optimal value of the functional of Eq. \ref{funct1}
\begin{eqnarray}
 {\cal L}(Q^*_.) = \sum_{v} Q^*_{v} 
\left[ \nu_{v}-    \ln \frac{Q^*_{v}  }{p_{v}   } \right] = 1+\eta = \ln  \left(  \sum_{v} p_{v} e^{ \nu_{v}  }   \right) =  \Phi [ \nu_{.}  ] 
\label{funct1opt}
\end{eqnarray}
indeed coincides with the result of Eq. \ref{phi1}.

\subsubsection{ Link with the relative entropy via the reciprocal Legendre transform }

The reciprocal Legendre transform of Eq. \ref{phi1srel}
reads
\begin{eqnarray}
S^{rel} (Q_{.}  \vert p_.  )   && 
=  \max \limits_{ \nu_{.}   } \left[ \sum_{v} \nu_{v} Q_{v} 
- \Phi [ \nu_{.}  ]  \right]
= \max \limits_{ \nu_{.}   } \left[ \sum_{v} \nu_{v} Q_{v} 
-  \ln  \left(  \sum_{v} p_{v} e^{ \nu_{v}  }   \right) \right]
\label{phi1recip}
\end{eqnarray}
The optimization over $\nu_{v}$ 
\begin{eqnarray}
0 && 
=\frac{ \partial  }{\partial \nu_{v} } \left[ \sum_{v'} \nu_{v'} Q_{v'} 
-  \ln  \left(  \sum_{v'} p_{v'} e^{ \nu_{v'}  }   \right) \right]
= Q_{v} -\frac{p_{v} e^{ \nu_{v}} }{\sum_{v'} p_{v'} e^{ \nu_{v'}  } }
\label{derirecip}
\end{eqnarray}
yields the optimal solution
\begin{eqnarray}
\nu^*_{v}
= \ln \left[ \frac{ Q_{v} \left( \sum_{v'} p_{v'} e^{ \nu_{v'}  } \right) }{p_{v} } \right]
\label{nuopt}
\end{eqnarray}
and the optimal value of the functional of Eq. \ref{phi1recip}
\begin{eqnarray}
 \max \limits_{ \nu_{.}   } \left[ \sum_{v} \nu_{v} Q_{v} 
-  \ln  \left(  \sum_{v} p_{v} e^{ \nu_{v}  }   \right) \right]
= \sum_{v} \nu^*_{v} Q_{v} 
-  \ln  \left(  \sum_{v} p_{v} e^{ \nu^*_{v}  }   \right)
= \sum_{v}  Q_{v} \ln \left[ \frac{ Q_{v}  }{p_{v} } \right] = S^{rel} (Q_{.}  \vert p_.  ) 
\label{maxopti}
\end{eqnarray}
coincides with the relative entropy as it should.

These calculations based on generating functions, Laplace's saddle-point method with constraints
taken into account via Lagrange multipliers, and Legendre transforms are very standard
both in statistical physics and in the theory of large deviations.

\subsection{ Third approach via some appropriate change of measure }

The third approach via some appropriate change of measure
is very common in the whole field of large deviations,
but appears to be less well known among physicists.
It seems thus useful to explain it here in more physical terms than usual.
The starting point is that the probability of the disorder configuration $[v(x)]_{x=1,2,..,L}$
of Eq. \ref{ptrajindep} can be rewritten only in terms of the empirical 1-point histogram of Eq. \ref{q1}
\begin{eqnarray}
 {\mathbb P}_L [v(.)]  = e^{\displaystyle  \sum_{x=1}^L \ln (p_{v(x)}) }
= e^{  \displaystyle L \sum_{v} Q_{v} \ln (p_{v})   }
\label{ptrajempi1}
\end{eqnarray}
So all the disorder configurations that have the same empirical 1-point histogram $Q_{. }  $ have the same probability
in Eq. \ref{ptrajempi1}.
As a consequence, the normalization of Eq. \ref{ptrajempi1}
over all disorder configurations $[v(.) ] $
can be rewritten 
 as a sum over the possible empirical 1-point histogram $Q_{.}$ 
\begin{eqnarray}
 1  = \sum_{ [v(.) ]}  {\mathbb P}_L [v(.)]
  =  \sum_{Q_. } C_1 [ Q_. ]  \Omega_L [  Q_{.} ] e^{ L \displaystyle \sum_{v}  Q_{v } \ln ( p_{v }  )  }  
\label{normaptrajchain}
\end{eqnarray}
where
\begin{eqnarray}
\Omega_L [Q_{.}  ] \equiv  \sum_{ v(.) }  \prod_{v} \left(
Q_{v }  - \frac{1}{L}  \sum_{x=1}^{L} \delta_{ v, v (x)} \right)
\label{omegaq1}
\end{eqnarray}
counts the number of disorder configurations that are associated to the same value $Q_.$
of the empirical histogram.
So the probability $P_L [ Q_{.}  ]  $ of Eq \ref{defprobaq1}
to observe the empirical histogram $Q_{.}$ reads
\begin{eqnarray}
P_L [ Q_{.}  ]    && =C_1[Q_.]  \Omega_L [  Q_{.} ] e^{ L \displaystyle \sum_{v}  Q_{v } \ln ( p_{v }  )  }    
\label{p1q1}
\end{eqnarray}

When the empirical 1-point histogram takes its typical value $p_v$ of Eq. \ref{q1typ},
the probability of Eq. \ref{p1q1} 
\begin{eqnarray}
P_L [ Q^{typ}_{.}  ]    && =P_L [ p_{.}  ] = \Omega_L [  p_{.} ] e^{ L \displaystyle \sum_{v}  p_{v } \ln ( p_{v }  )  }    
\label{p1q1typ}
\end{eqnarray}
should not decay exponentially in $L$,
so that $ \Omega_L [  p_{.} ]$ should grow exponentially in $L$ in order to compensate exactly the other exponential factor
\begin{eqnarray}
 \Omega_L [  p_{.} ] \opsimeq_{L \to +\infty} e^{ - L \displaystyle \sum_{v}  p_{v } \ln ( p_{v }  )  }    
\label{omegap1}
\end{eqnarray}

To obtain the behavior of $\Omega_L [  Q_{.} ] $ when the empirical 1-point histogram $Q_{.} $
is different from its typical value $Q_{.}^{typ} = p_{. }   $,
we may consider a modified model where the disorder is drawn with the modified
probability ${\tilde p}_{v}=Q_{v}$ that will make $Q_{v}$ typical for this modified model,
and one obtains
\begin{eqnarray}
\Omega_L [Q_{.}  ] \opsimeq_{L \to +\infty}   \ e^{ L S_1 [Q_{.}  ]  }  
\label{omega1}
\end{eqnarray}
where 
\begin{eqnarray}
S_1 [Q_{.}  ]  \equiv  - \sum_{v} Q_{v} \ln \left(   Q_{v}\right)
\label{entropy1}
\end{eqnarray}
 represents the entropy of the empirical 1-point histogram $Q_{.}$.
Plugging this result into Eq. \ref{p1q1} yields that the large deviation behavior
of the probability of the empirical 1-point histogram
\begin{eqnarray}
P_L [ Q_{.}  ]    && =C_1 [ Q_{.} ]\Omega_L [  Q_{.} ] e^{  \displaystyle L \sum_{v}  Q_{v } \ln ( p_{v }  )  }   
\opsimeq_{L \to +\infty} 
C_1 [ Q_{.} ]   \ e^{-L   \sum_{v} Q_{v} \ln \left(  \frac{ Q_{v} }{ p_{v } } \right)    }  
= C_1 [ Q_{.} ]   \ e^{ L S^{rel} (Q_{.}  \vert p_.  ) }
\label{p1q1third}
\end{eqnarray}
involves again the relative entropy $S^{rel} (Q_{.}  \vert p_.  ) $ as it should to recover Eq \ref{sanov} and Eq. \ref{s1relative}.

This idea to evaluate the large deviations properties of the untypical values of the empirical observable
via the introduction of a modified model that make this empirical observable typical
is used extensively in the field of large deviation for the two following reasons.
From the conceptual point of view, this way of thinking is very illuminating 
because it shows very clearly why the entropy $S_1 [Q_{.}  ] $ appears in Eq. \ref{omega1}
 and why the relative entropy $S^{rel} (Q_{.}  \vert p_.  ) $ appears in Eq \ref{p1q1third}.
From the technical point of view, it is extremely powerful, since it allows to obtain
directly the results without any actual computations : indeed, one does not need 
to use combinatorics to enumerate the appropriate configurations in finite size as in Eq. \ref{multinomial},
and one does not need either to compute the generating function of Eq. \ref{geneq1} and to perform 
the reciprocal Legendre transform, but one obtains directly the rate function from simple considerations.
 In the following sections concerning the more complicated cases of empirical histograms of higher orders, 
as well as in the Appendix, we will see how this approach can be adapted to each purpose
 in order to obtain directly the appropriate rate functions without any calculation.


\section{ Empirical 2-point histogram as the Level 2.5 of large deviations }

\label{sec_q2}

In this section, we focus on the probability of the empirical 2-point histogram of Eq. \ref{qr} for $r=2$
over the disorder configurations $[v(.)]$ drawn with Eq. \ref{ptrajindep}
\begin{eqnarray}
P_L [ Q_{..}  ]    \equiv \sum_{[v(.)] }  {\mathbb P}_L[v(.)] \prod_{v_2} \prod_{v_1} 
\delta \left(Q_{v_2 v_1 } - \frac{1}{L}  \sum_{x=1}^{L} \delta_{ v_2, v (x+2 )}\delta_{ v_1, v (x+1 )}   \right)
\label{defprobaq2}
\end{eqnarray}
Its large deviations properties have been analyzed in the context of Markov chains \cite{fortelle_thesis,fortelle_chain,review_Touchette}. 
Together with its analog formulations for Markov jump processes in continuous time \cite{fortelle_thesis,fortelle_jump,maes_canonical,maes_onandbeyond,wynants_thesis,chetrite_formal,BFG1,BFG2,chetrite_HDR,LargeDevRings,largeDevOpen,mFT}
and for diffusion processes \cite{wynants_thesis,maes_diffusion,chetrite_formal,engel,chetrite_HDR},
it is nowadays called the 'Level 2.5' in the field of large deviations.

\subsection{ Constraints on the  empirical 2-point histogram $Q_{..}$ }

Since the empirical 1-point histogram $Q_.$ can be reconstructed by summing over the last or the first value
of the empirical 2-point histogram $Q_{..}$, 
it is convenient to introduce the following notation to summarize these constraints
\begin{eqnarray}
C_2 [ Q_{..},Q_{.} ] = \prod_{v_1} \left[ \delta \left( Q_{v_1 }  - \sum_{v_2} Q_{v_2 v_1}  \right)
\delta \left( Q_{v_1 } - \sum_{v_2} Q_{v_1 v_2} \right) \right]
\label{c2}
\end{eqnarray}
while the empirical 1-point histogram $Q_.$ should of course 
still satisfy the normalization constraint $C_1[Q_.]$ of Eq. \ref{c1}.

\subsection{ Generalized Markovian model for the disorder }

In order to analyze the statistical properties of the empirical 2-point histogram,
it is useful to introduce a generalized model where the disorder configurations are
generated by a Markov chain where the transition probability matrix $W_{v' v} $
to go from $v$ to $v'$ is normalized to unity
\begin{eqnarray}
\sum_{v'} W_{v' v} && = 1
\label{wanorma}
\end{eqnarray}
The probability of Eq. \ref{ptrajindep} for a disorder configuration $[v(x)]_{x=1,2,..,L} $
is thus replaced by the product of the transition probabilities along the configuration (up to boundary terms that become negligible 
for large $L \to +\infty$)
\begin{eqnarray}
{\mathbb P}^{Markov}_L [v(.)] 
  \simeq \prod_{x=1}^{L}  W_{v(x+1) v(x)}
= e^{ \displaystyle L \sum_{x=1}^{L}  \ln \left( W_{v(x+1) v(x)} \right) }
\label{ptrajmarkova}
\end{eqnarray}

It is also useful to introduce the stationary state $\rho_{v}$ of this Markov chain
satisfying 
\begin{eqnarray}
\rho_{v'} && = \sum_{v} W_{v' v} \rho_{v}
\label{rhoa}
\end{eqnarray}
 with the normalization 
\begin{eqnarray}
1=  \sum_{v} \rho_{v}
\label{normarhoa}
\end{eqnarray}

For this generalized Markovian model, the typical value of the empirical 1-point histogram
of Eq. \ref{q1} is simply the stationary state $\rho_{v} $ introduced in Eq. \ref{rhoa}
\begin{eqnarray}
Q^{typ}_{v} && = \rho_{v}
\label{q1typgen}
\end{eqnarray}
while the typical value of the empirical 2-point histogram 
 is given by the corresponding flow appearing in Eq. \ref{rhoa}
\begin{eqnarray}
Q^{typ}_{v' v} = W_{v' v} \rho_{v}
\label{q2typgen}
\end{eqnarray}
that satisfy the constraints of Eqs \ref{c2} and Eq \ref{c1}.

Since the probability of Eq. \ref{ptrajmarkova}
can be rewritten only in terms of the empirical 2-point histogram $Q_{v' v} $ as
\begin{eqnarray}
P^{Markov}_L [v(.)]   \simeq 
 e^{ \displaystyle L \sum_{v'}\sum_{v} Q_{v' v} \ln \left( W_{v' v} \right) }
\label{ptrajmarkovaempi}
\end{eqnarray}
the normalization over disorder configurations can be rewritten as a sum over the empirical 1-point and 2-point 
histograms with their constraints of Eq. \ref{c1} and Eq \ref{c2} as
\begin{eqnarray}
1= \sum_{v(. )} P^{Markov}_L  [v(. )]
=  \sum_{Q_. } C_1[Q_.] \sum_{Q_{..} } C_2[Q_{..},Q_.] \ \Omega_L [  Q_{..},Q_. ] 
e^{ \displaystyle L \sum_{v'}\sum_{v} Q_{v' v} \ln \left( W_{v' v} \right)  }  
\label{ptrajmarkovaempin}
\end{eqnarray}
where $\Omega_L [  Q_{..},Q_. ]  $ counts the number of disorder configurations that have the empirical observables $[ Q_{..},Q_. ]$ and is thus the direct generalization of Eq. \ref{omegaq1}, 
while the probability to observe these empirical observables reads
\begin{eqnarray}
P_L[ Q_{..},Q_. ] = C_1[Q_.]  C_2[Q_{..},Q_.] \ \Omega_L [  Q_{..},Q_. ] 
e^{ \displaystyle L \sum_{v'}\sum_{v} Q_{v' v} \ln \left( W_{v' v} \right)  }  
\label{pempi2}
\end{eqnarray}
For the typical values of Eq. \ref{q1typgen} and Eq \ref{q2typgen} of the empirical observables,
this probability should not be exponentially small in $L$ so that $\Omega_L [  Q^{typ}_{..},Q^{typ}_. ]  $
should exactly compensate the other exponential factor in Eq. \ref{pempi2}
\begin{eqnarray}
 \Omega_L [  Q^{typ}_{..},Q^{typ}_. ] \opsimeq_{L \to +\infty} 
e^{ - \displaystyle L \sum_{v'}\sum_{v} Q^{typ}_{v' v} \ln \left( W_{v' v} \right)  }  
\label{omega2typ}
\end{eqnarray}

For other values of the empirical observables, one may consider a modified Markov transition matrix
${\tilde W}_{v' v} $ that would make these empirical histograms typical : Eqs \ref{q1typgen} and \ref{q2typgen}
yields that the appropriate choice is
\begin{eqnarray}
{\tilde W}_{v' v}= \frac{Q_{v' v}}{Q_{v} } 
\label{wamod}
\end{eqnarray}
so that Eq \ref{omega2typ} becomes
\begin{eqnarray}
 \Omega_L [  Q_{..},Q_. ] \opsimeq_{L \to +\infty} 
e^{ - \displaystyle L \sum_{v'}\sum_{v} Q_{v' v} \ln \left( {\tilde W}_{v' v} \right)  }  
= e^{ - \displaystyle L \sum_{v'}\sum_{v} Q_{v' v} \ln \left( \frac{Q_{v' v}}{Q_{v} }  \right)  }  
\opsimeq_{L \to +\infty} 
 e^{  \displaystyle L \left(S_2 [Q_{..}  ]  - S_1 [Q_{.}  ]   \right) }  
\label{omega2}
\end{eqnarray}
where $S_2 [Q_{..}  ] $ represents the entropy of the empirical 2-point histogram $Q_{..}$
\begin{eqnarray}
S_2 [Q_{..}  ]  \equiv  - \sum_{v'} \sum_{v} Q_{v' v} \ln \left(   Q_{v' v}\right)
\label{entropy2}
\end{eqnarray}
while $S_1[Q_.]$ is the entropy of the empirical 1-point histogram $Q_{.}$ introduced in Eq \ref{entropy1}.

Plugging Eq. \ref{omega2} into Eq \ref{pempi2}
yields the large deviation form \cite{fortelle_thesis,fortelle_chain,review_Touchette}
\begin{eqnarray}
P_L[ Q_{..},Q_. ] \opsimeq_{L \to +\infty}  C_1[Q_.]  C_2[Q_{..},Q_.] 
 e^{ - \displaystyle L \sum_{v'}\sum_{v} Q_{v' v} \ln \left( \frac{Q_{v' v}}{ W_{v' v}Q_{v} }  \right)  } 
\label{pempi2.5}
\end{eqnarray}
that is called nowadays the 'Level 2.5' for Markov chains.
The rate function can be interpreted as the relative entropy for Markov chains \cite{fortelle_thesis,fortelle_chain,review_Touchette}.
The analog results have been much studied for Markov jump processes in continuous time \cite{fortelle_thesis,fortelle_jump,maes_canonical,maes_onandbeyond,wynants_thesis,chetrite_formal,BFG1,BFG2,chetrite_HDR,LargeDevRings,largeDevOpen,mFT}
and for diffusion processes \cite{wynants_thesis,maes_diffusion,chetrite_formal,engel,chetrite_HDR},

\subsection{ Return to the initial disorder of Eq. \ref{ptrajindep}  }

The initial disorder model of Eq. \ref{ptrajindep} corresponds to the special case
where the Markov matrix of Eq. \ref{wanorma} reduces to
\begin{eqnarray}
W_{v' v} = p_{v'}
\label{passpecialW}
\end{eqnarray}
Then Eq \ref{pempi2.5} simplifies into
\begin{eqnarray}
P_L[ Q_{..},Q_. ]  \opsimeq_{L \to +\infty}  
C_1[Q_.]  C_2[Q_{..},Q_.] 
 e^{ - \displaystyle L \sum_{v'}\sum_{v} Q_{v' v} \ln \left( \frac{Q_{v' v}}{ p_{v' }Q_{v} }  \right)  } 
 = C_1 [ Q_{.} ]   \ e^{ L S^{rel} (Q_{.}  \vert p_.  ) } C_2[Q_{..},Q_.]  \ e^{T \left( S_2 [Q_{..}  ]  - 2S_1 [Q_{.} ] \right) }
\label{pempi2final}
\end{eqnarray}
In the last expression, one recognizes the probability $P_L [ Q_{.}  ]  $ of the empirical 1-point histogram $Q_.$
of Eq. \ref{p1q1third}. This yields that the conditional probability to observe the 
empirical 2-point histogram $Q_{..}$ once the empirical 1-point histogram $Q_.$ is given reads
\begin{eqnarray}
 P_L [ Q_{..} \vert Q_. ]     = \frac{P_L [ Q_{..} , Q_. ]  } {P_L [ Q_{.}  ] }
&&  \opsimeq_{L \to +\infty} 
 C_2 [ Q_{..},Q_{.} ] 
e^{ L \left( S_2 [Q_{..}  ] - 2 S_1 [Q_{.}  ]  \right) } 
\label{p2q2deb}
\end{eqnarray}
In particular, once the empirical 1-point histogram $Q_.$ is given,
the typical value of the empirical 2-point histogram $Q_{..}$ is simply the product
\begin{eqnarray}
 Q^{typ} _{v_2 v_1}  = Q_{v_2 } Q_{v_1}
\label{q2typ}
\end{eqnarray}
as it should, while Eq \ref{p2q2deb} described the large deviations away from this typical value.

\section{ Empirical higher order histograms as the level 3 of large deviations }

\label{sec_level3}

In the language of large deviations, the Level 3 actually denotes the empirical process 
that can be constructed from the knowledge of the empirical r-point histogram in the limit $r \to +\infty$ \cite{oono,review_Touchette}.
In this section, we will not be interested into taking this limit, 
but we wish to analyze the hierarchy of the empirical r-point histograms of arbitrary order $r$
up to the maximal value $r_{max}=L$ (Eq \ref{qrmax}), in order to characterize the sample-to-sample fluctuations for a disordered ring of large size $L$. So strictly speaking, this section is between the Level 2.5 of the previous section
and the Level 3 concerning the limit $r \to +\infty$.

\subsection{ Large deviations properties of the empirical r-point histograms of arbitrary order $r$ }

In the two previous sections, we have described in detail the 
large deviations properties of the empirical 1-point histogram $Q_.$ and 2-point histogram $Q_{..}$.
Via iteration, one may analyze similarly the properties of the empirical r-point histogram $Q_{(rpoints)}$ of Eq. \ref{qr}
of arbitrary order $r$. Since the empirical $(r-1)$-point histogram $Q_{((r-1)points)}$
 can be reconstructed by summing over the last or the first value of the r-point histogram $Q_{(rpoints)}$,
it is convenient to introduce the following notation analogous to Eq \ref{c2} to summarize them
\begin{eqnarray}
C_r [ Q_{(rpoints)} ,Q_{((r-1)points)} ] =\prod_{v_{r-1}} ... \prod_{v_1}
\delta \left(  Q_{v_{r-1}  ... v_1}  - \sum_{v_r} Q_{v_r ... v_1} \right)
\delta \left(  Q_{v_{r-1}  ... v_1}  - \sum_{v_r} Q_{v_{r-1}  ... v_1 v_r}  \right)
\label{ck}
\end{eqnarray}

The final result is that
 the probability $P_L(Q_{(rpoints)},...,Q_{..},Q_.) $ to observe the 
empirical histograms up to the $r$-point histogram $Q_{(rpoints)} $
normalized to unity
\begin{eqnarray}
1 && = \sum_{Q_{(rpoints)} } ... \sum_{Q_{..} } \sum_{Q_{.} } P_L(Q_{(rpoints)},...,Q_{..},Q_.)
\label{normak}
\end{eqnarray}
follows the large deviation form
\begin{eqnarray}
P_L(Q_{(rpoints)},...,Q_.)
 \opsimeq_{L \to +\infty}  &&
  \left( C_1 [ Q_{.} ]   e^{ L \displaystyle \sum_{v}  Q_{v } \ln ( p_{v } ) } \right)
  C_2 [ Q_{..} ,Q_{.}] 
C_3 [ Q_{...} ,Q_{..}]
...
 C_{r-1} [ Q_{((r-1)points)}, Q_{((r-2)points)} ] 
\nonumber \\
&&
C_r [ Q_{(rpoints)} , Q_{((r-1)points)}] 
 e^{ L \displaystyle \left( S_r [Q_{(rpoints)}  ] - S_{r-1} [Q_{((r-1)points)}  ]  \right) } 
\label{ptkinter}
\end{eqnarray}
that generalizes Eq. \ref{pempi2final}.
Besides the consistency constraints $(C_1,..,C_r)$ up to order $r$ (Eq \ref{ck})
and besides the disorder configuration weight $e^{ L  \sum_{v}  Q_{v } \ln ( p_{v } ) }  $
of Eq. \ref{ptrajempi1} that only involves the empirical 1-point histogram $Q_{v }$,
the remaining factor
corresponds to the exponential growth of the number of configurations that have some 
empirical r-point histogram $Q_{(rpoints)}$
\begin{eqnarray}
\Omega_L [ Q_{(rpoints)} ; Q_{((r-1)points)} ]
 \opsimeq_{L \to +\infty}   e^{ L \left( S_r [Q_{(rpoints)}  ] - S_{r-1} [Q_{((r-1)points)}  ]  \right) } 
\label{omega0k}
\end{eqnarray}
 in terms of
the entropy of the empirical r-point histogram $Q_{(rpoints)}$
\begin{eqnarray}
S_r [Q_{(rpoints)}  ]  \equiv  - \sum_{v_r} ... \sum_{v_2} \sum_{v_1}
Q_{v_r ...v_2 v_1} \ln \left(   Q_{v_r ...v_2 v_1}\right)
\label{entropyk}
\end{eqnarray}

Equivalently, Eq. \ref{ptkinter}
means that the conditional probability 
to observe the empirical $r$-point histogram $ Q_{(rpoints)}$ once the 
empirical $(r-1)$-point histogram $ Q_{((r-1)points)}$ is given
reads
\begin{eqnarray}
 && P_L [ Q_{(rpoints)} \vert Q_{((r-1)points)} ]    
 \equiv \frac{P_L(Q_{(rpoints)},...,Q_{..},Q_.)}{P_L(Q_{((r-1)points)},...,Q_{..},Q_.)}
\nonumber \\
&& \opsimeq_{L \to +\infty} 
 C_r [  Q_{(rpoints)} , Q_{((r-1)points)}  ] 
e^{ L \left( S_r [Q_{(rpoints)}  ] - 2 S_{r-1} [Q_{((r-1)points)} ]+S_{r-2} [Q_{((r-2)points)}    ]  \right) } 
\label{pkconditional}
\end{eqnarray}
which is the generalization of Eq. \ref{p2q2deb}.

\subsection{ Analysis of the hierarchy in the backward direction via contraction }

Up to now we have described the hierarchy of empirical histograms 
by considering successively higher and higher order $r$.
But it is also useful to see now how one goes backwards in this hierarchy,
via the notion of 'contraction' which is the generic name in the field of large deviations
for the operation needed to go from a higher to a lower level of description.
In our present case, the contraction consists in finding the optimal empirical $r$-point histogram that maximizes 
the conditional probability 
Eq \ref{pkconditional} when 
all the lower-order empirical histograms are given.
 One needs to maximize 
the exponential factor in Eq \ref{pkconditional}
in the presence of the constraints $C_r [ Q_{(rpoints)}, Q_{(r-1)points} ]   $ of Eq. \ref{ck}
that can be taken into account via Lagrange multipliers.
So one considers the following functional of $ Q_{(rpoints)} $
\begin{eqnarray}
&&  {\cal L} [ Q_{(rpoints)}  ]  =  
- \sum_{v_r} ...\sum_{v_2} \sum_{v_1}
Q_{v_r ...v_2 v_1} \ln \left(   Q_{v_r ...v_2 v_1}\right)
- 2 S_{r-1} [Q_{((r-1)points)}] +S_{r-2} [Q_{((r-2)points)}]
 \nonumber \\
&&+ \sum_{v_{r-1}} ...   \sum_{v_2} \sum_{v_1} 
g_{v_{r-1}  ... v_1  }
\left(  
Q_{v_{r-1}  ... v_1}  - \sum_{v_r} Q_{v_r ... v_1} 
\right) 
+ \sum_{v_{r}}  ... \sum_{v_3} \sum_{v_2} 
 f_{v_{r}  ... v_2  }
 \left( 
Q_{v_{r}  ... v_2}  - \sum_{v_1} Q_{v_{r}  ... v_2 v_1}  
\right)
\label{lk}
\end{eqnarray}
The optimization with respect to $Q_{v_r ...v_2 v_1}  $
\begin{eqnarray}
0 = \frac{ \partial  {\cal L} [ Q_{(rpoints)}  ] }{ \partial Q_{v_r ...v_2 v_1} } 
 =   -  \ln \left(   Q_{v_r ...v_2 v_1}\right) -1 
-  g_{v_{r-1}  ... v_1  }
-  f_{v_{r}  ... v_2  }
\label{lkderi}
\end{eqnarray}
yields the optimal solution
\begin{eqnarray}
  Q_{v_r ...v_2 v_1}^{*} =
 e^{ -1 - f_{v_{r}  ... v_2  }  - g_{v_{r-1}  ... v_1  } }
\label{lkderisol}
\end{eqnarray}
where the Lagrange multipliers $ f_{v_{r}  ... v_2  }$ and $g_{v_{r-1}  ... v_1  } $ have to be chosen to satisfy the constraints
\begin{eqnarray}
Q_{v_{r-1}  ... v_1}  && = \sum_{v_r} Q_{v_r ... v_1}^*
= e^{ -1} \left( \sum_{v_r} e^{ - f_{v_{r}  ... v_2  }  } \right)  e^{- g_{v_{r-1}  ... v_1  } }
\nonumber \\
Q_{v_{r}  ... v_2}  && =  \sum_{v_1} Q_{v_{r}  ... v_2 v_1}^*
= e^{ -1} e^{ - f_{v_{r}  ... v_2  }  } \left( \sum_{v_1}   e^{- g_{v_{r-1}  ... v_1  } } \right) 
\label{lkc}
\end{eqnarray}
A further consequence is thus the following constraint involving the empirical histogram of order $(k-2)$
\begin{eqnarray}
Q_{v_{r-1}  ... v_2}  && = \sum_{v_r} \sum_{v_1} Q_{v_r ... v_1}^*
= e^{ -1} \left( \sum_{v_r} e^{ - f_{v_{r}  ... v_2  }  } \right)   \left( \sum_{v_1}   e^{- g_{v_{r-1}  ... v_1  } } \right) 
\label{dericonstraint}
\end{eqnarray}

These four last equations yield that the optimal solution of Eq. \ref{lkderisol} 
can be simply rewritten
as the product of the two empirical observables of order $(k-1)$ of Eq \ref{lkc} 
divided by the empirical observable of order $(k-2)$ of  Eq \ref{dericonstraint}
\begin{eqnarray}
  Q_{v_r ...v_2 v_1}^* = \frac{Q_{v_{r}  ... v_2} Q_{v_{r-1}  ... v_1}}{Q_{v_{r-1}  ... v_2}}
\label{qksoluratio}
\end{eqnarray}

One then needs to evaluate the entropy of Eq. \ref{entropyk} of
 this optimal solution $ Q_{v_r ...v_2 v_1}^*$
\begin{eqnarray}
&& S_r [Q_{(rpoints)}^*  ]   =  - \sum_{v_r} ... \sum_{v_2} \sum_{v_1}
Q_{v_r ...v_2 v_1}^* \ln \left(   Q_{v_r ...v_2 v_1}^* \right)
\nonumber \\
&& =   - \sum_{v_r} ... \sum_{v_2} \sum_{v_1}
 \frac{Q_{v_{r}  ... v_2} Q_{v_{r-1}  ... v_1}}{Q_{v_{r-1}  ... v_2}}
\left[ \ln ( Q_{v_{r}  ... v_2}) + \ln (Q_{v_{r-1}  ... v_1} ) - \ln (Q_{v_{r-1}  ... v_2} )  \right]
\nonumber \\
&& =   - \sum_{v_r} ... \sum_{v_2} 
 Q_{v_{r}  ... v_2} 
 \ln ( Q_{v_{r}  ... v_2}) 
 - \sum_{v_{r-1}} ... \sum_{v_2} \sum_{v_1}
  Q_{v_{r-1}  ... v_1}
 \ln (Q_{v_{r-1}  ... v_1} ) 
 + \sum_{v_{r-1}} ... \sum_{v_2} 
 Q_{v_{r-1}  ... v_2}
 \ln (Q_{v_{r-1}  ... v_2} )  
\nonumber \\
&& =   2 S_{r-1)} [Q_{((r-1)points)}] - S_{r-2)} [Q_{((r-2)points)}]
\label{entropykopt}
\end{eqnarray}
So the functional of Eq. \ref{lk} vanishes for this optimal solution $Q^*_{(rpoints)} $
\begin{eqnarray}
&&  {\cal L} [ Q^*_{(rpoints)}  ]  =  0
\label{lkopt}
\end{eqnarray}
i.e. the conditional probability of Eq. \ref{pkconditional} does not decay exponentially in $L$
for this optimal solution $Q^*_{(rpoints)}  $, that represents the typical value of $Q_{v_r ...v_2 v_1} $
once all the empirical histograms of lower order are given 
\begin{eqnarray}
  Q_{v_r ...v_2 v_1}^{typ} =  Q_{v_r ...v_2 v_1}^* = \frac{Q_{v_{r}  ... v_2} Q_{v_{r-1}  ... v_1}}{Q_{v_{r-1}  ... v_2}}
\label{qksoluratiotyp}
\end{eqnarray}
The probability of all other values is described by the large deviation form of Eq. \ref{pkconditional}.

\section{ Random models on the Cayley tree from large deviations of branches }

\label{sec_tree}


Many random models have been studied on the geometry of the Cayley tree,
where the absence of loops allows to write exact recurrences on probability distributions :
two famous examples are the Directed Polymer on the Cayley tree \cite{Der_Spo,Coo_Der} 
and the Anderson Localization on the Cayley tree \cite{abou1,abou2,DR,MD}.
In the Cayley tree of branching ratio $K$ around the central root $O$,
the number of sites at distance $r$ 
\begin{eqnarray}
N^{tree}(r) = (K+1) K^{r-1}  =  \frac{K+1}{K} K^r
\label{nrcayley}
\end{eqnarray}
grows exponentially with the distance $r$, in contrast to the power-law growth as $r^{d-1}$ in any finite dimension $d$.
The Cayley tree is thus considered as an appropriate way to define the mean-field version of
 random models in infinite dimensionality $d=\infty$.

It is interesting to compare the properties of the same random model defined in the two following geometries :

(i) in the finite Cayley tree of branching ratio $K$ with $L$ generations around the central root $O$,
where the number of leaves is given by Eq. \ref{nrcayley} for $r=L$
\begin{eqnarray}
N^{tree}(L) =   \frac{K+1}{K} K^L
\label{nlcayley}
\end{eqnarray}

(ii) in the star geometry, where the central root $O$ is linked to $K^L$ independent one-dimensional lattices of $L$ sites,
so that the number of sites at distance $r$ is actually independent of $r$
\begin{eqnarray}
N^{star}_{K^L}(r) = K^L 
\label{nlstar}
\end{eqnarray}
but the number of leaves at $r=L$ displays the same exponential behavior in $L$ as Eq. \ref{nlcayley}.

Although (ii) may look as an extremely crude approximation of (i), 
the properties of some random models defined on (i) and (ii) have turned out to be very close,
as exemplified by the exact solutions of (i) the Directed Polymer on the Cayley tree \cite{Der_Spo,Coo_Der}
and of (ii) the Directed Polymer in the star geometry that coincides with the Random Energy Model \cite{rem}
(a model that had been introduced before with completely different motivations coming from mean-field spin-glasses). The differences between the two only appear in the finite-size scaling properties of the freezing transition \cite{Coo_Der}.

In the star geometry (ii), it is clear that the random model 
will be governed by the large deviations properties
of the corresponding one-dimensional model of length $L$ 
that appear on the $K^L$ independent branches.
In this section, the goal is thus to describe 
how the large deviations properties
of one-dimensional models that have been discussed in the previous sections
can be used to analyze the properties of the same model on this star geometry (ii).

\subsection { Model on the star geometry where each branch corresponds to a product of random variables }

\label{sec_star}

We wish the analyze the star geometry (ii) above,
where each of the independent $K^L$ branches labelled by $b=1,2,..,K^L$
can be described by a product of $L$ random variables as Eq. \ref{prodrandom}
\begin{eqnarray}
  \tau_L [v_b(. )] = \prod_{x=1}^L  t_{v_b(x)}
\label{prodrandomb}
\end{eqnarray}
with its corresponding finite-size Lyapunov exponent of Eq. \ref{lyapunovprod}
\begin{eqnarray}
\lambda_b \equiv \lambda [v_b(. )]   \equiv \frac{ \ln  \vert \tau_L [v(. )]  \vert  }{ L} = \frac{1}{L} \sum_{x=1}^L  \ln \vert t_{v_b(x)} \vert
\label{lyapunovprodb}
\end{eqnarray}
whose large deviations properties for large $L$ are described by some rate function $I(\lambda)$ 
\begin{eqnarray}
{\cal P}_L(\lambda)  \opsimeq_{ L \to + \infty} e^{- L I(\lambda) }
\label{largedevlyapunovb}
\end{eqnarray}

Each disordered configuration on the star geometry
can be then characterized  by the empirical histogram of the Lyapunov exponent $\lambda_b$ of Eq. \ref{lyapunovprodb}
for the $K^L$ independent branches
\begin{eqnarray}
{\cal Q}_{L} (\lambda) \equiv   \frac{1}{K^L} \sum_{b=1}^{K^L} \delta( \lambda-\lambda_b)
\label{qempib}
\end{eqnarray}
while the empirical number of branches having the Lyapunov exponent $\lambda$ reads
\begin{eqnarray}
{\cal N}_{L} (\lambda) \equiv \sum_{b=1}^{K^L} \delta( \lambda-\lambda_b) = K^L {\cal Q}_{L} (\lambda)
\label{nempib}
\end{eqnarray}

In various models, an interesting class of observables are
 given by the sums over the $K^L$ independent branches
of the powers of non-integer $k$ of the products $\tau_L [v_b(. )] $ of Eq. \ref{prodrandomb}
\begin{eqnarray}
{\cal S}_{L}(k) \equiv \sum_{b=1}^{K^L}   \vert \tau_L [v_b(. )] \vert^k
\label{calmkdef}
\end{eqnarray}
that can be rewritten in terms of the Lyapunov exponents $\lambda_b$ (Eq \ref{lyapunovprodb})
of the $K^L$ branches or in terms of the empirical observables of Eqs \ref{qempib} and \ref{nempib} as
\begin{eqnarray}
{\cal S}_{L}(k) 
= \sum_{b=1}^{K^L} e^{ k L \lambda_b} = 
\int d\lambda e^{ k L \lambda} {\cal N}_L(\lambda) = K^L \int d\lambda e^{ k L \lambda} {\cal Q}_{L} (\lambda)
\label{calmk}
\end{eqnarray}

\subsection { Statistical properties of the empirical histogram ${\cal Q}_{L} (\lambda) $ of the Lyapunov exponent }

The typical value of the empirical histogram of Eq. \ref{qempib}
is given by the true probability of the Lyapunov exponent of Eq. \ref{largedevlyapunovb}
\begin{eqnarray}
{\cal Q}^{typ}_{L} (\lambda) =   {\cal P}_L(\lambda)  \opsimeq_{ L \to + \infty} e^{- L I(\lambda) }
\label{empibtyp}
\end{eqnarray}
so that in a given sample, the empirical number of branches of Eq. \ref{nempib} has for typical value
\begin{eqnarray}
{\cal N}^{typ}_{L} (\lambda)  = K^L {\cal Q}_{L} (\lambda)\opsimeq_{ L \to + \infty} e^{ L \left[\ln K -  I(\lambda) \right] }
\label{nempibtyp}
\end{eqnarray}
The typical value $\lambda^{typ} $ of the one-dimensional model corresponding to the vanishing of the rate function
$ I(\lambda^{typ}) =0$ will thus appear in an extensive number of the branches
\begin{eqnarray}
{\cal N}^{typ}_{L} (\lambda^{typ})  \opsimeq_{ L \to + \infty} e^{ L \ln K  } =K^L
\label{nempibtypltyp}
\end{eqnarray}
while all the other values in the interval $\lambda^- < \lambda < \lambda^+ $ where
\begin{eqnarray}
I(\lambda) < \ln K = I(\lambda^- ) = I (\lambda^+)
\label{subextensive}
\end{eqnarray}
will appear in a sub-extensive number $e^{ L \left[\ln K -  I(\lambda) \right] } $ of branches.
Finally, the values of the Lyapunov exponent outside this interval, i.e. in the two regions $\lambda < \lambda^-$
and $\lambda > \lambda^+$ where the rate function satisfies $I(\lambda) > \ln K$
are too rare to appear in a typical sample of the star geometry, so that Eq. \ref{nempibtyp} should be rewritten more precisely 
for a typical sample as
\begin{eqnarray}
{\cal N}^{TypicalSample}_{L} (\lambda)  \opsimeq_{ L \to + \infty} e^{ L \left[\ln K -  I(\lambda) \right] } \theta( I(\lambda) \leq \ln K)
=  e^{ L \left[\ln K -  I(\lambda) \right] } \theta( \lambda^- \leq \lambda \leq \lambda^+  )
\label{nempibtypsample}
\end{eqnarray}

However the values $\lambda < \lambda^-$ and $\lambda > \lambda^+$ that do not appear in a typical sample
may appear in atypical samples, and it is thus interesting to consider the large deviations 
of the empirical histogram ${\cal Q}_{L} (.) $ of Eq. \ref{qempib}
with respect to its typical value ${\cal Q}^{typ}_{L} (.)= {\cal P}_L(.) $ of Eq. \ref{empibtyp} :
since the $K^L$ branches are independent, one may directly adapt the Sanov result of Eq. \ref{sanov}
to our present notations : the probability to observe the empirical histogram ${\cal Q}_{L} (.) $
follows the large deviation form with respect to the size $K^L$
\begin{eqnarray}
P_L [ {\cal Q}_L(.) ]   \opsimeq_{L \to +\infty} \delta \left( 1-  \int d\lambda  {\cal Q}_{L} (\lambda) \right) 
e^{- K^L S^{rel} ( {\cal Q}_{L}(.)  \vert {\cal P}_L(.)  )  }
\label{sanovtree}
\end{eqnarray}
where the rate function corresponds to the relative entropy
\begin{eqnarray}
 S^{rel} ( {\cal Q}_{L}(.)  \vert {\cal P}_L(.)  )  = \int d\lambda  {\cal Q}_{L} (\lambda) 
 \ln \left( \frac{ {\cal Q}_{L} (\lambda)  }{  {\cal P}_L( \lambda )   }  \right)
\label{s1relativetree}
\end{eqnarray}
of the empirical histogram ${\cal Q}_{L} (.) $
with respect to the true probability distribution ${\cal P}_L(.) $ of the Lyapunov exponent (Eq. \ref{largedevlyapunovb}).
As explained in detail in section \ref{sec_genelevel2},
the Sanov result of Eq. \ref{sanovtree}
is equivalent to the following expression of the generating function
that is valid for any finite $L$
 (Eq. \ref{geneq1} as adapted to our present context)
\begin{eqnarray}
 {\cal G}_L [ \nu(.)  ]  && 
\equiv  \sum_{ {\cal Q}_L(.) } P_L [ {\cal Q}_L(.)  ] 
 \ e^{ \displaystyle  K^L \int d\lambda  \nu(\lambda)  {\cal Q}_{L} (\lambda) }   
 =  \int d\lambda_1 ... \int d \lambda_{K^L} 
 {\cal P}_L( \lambda_1 ) ...  {\cal P}_L( \lambda_{K^L} ) 
 \  e^{ \displaystyle   \sum_{b=1}^{K^L} \nu(\lambda_b ) }   
 \nonumber \\ &&
 = \prod_{b=1}^{K^L}\left( \int d\lambda_b {\cal P}_L( \lambda_b )  e^{\nu(\lambda_b )}    \right)
= \left(  \int d\lambda {\cal P}_L( \lambda )  e^{\nu(\lambda )}     \right)^{K^L}  
\label{geneq1tree}
\end{eqnarray}

In particular, the successive derivatives with respect to $\nu(\lambda)$
\begin{eqnarray}
\frac{\partial   {\cal G}_L [ \nu(.)  ] }{\partial \nu(\lambda ) }   && 
=  \sum_{ {\cal Q}_L(.) } P_L [ {\cal Q}_L(.)  ] \ K^L  {\cal Q}_{L} (\lambda)
 \ e^{ \displaystyle  K^L \int d\lambda'  \nu(\lambda')  {\cal Q}_{L} (\lambda') }   
= K^L  {\cal P}_L( \lambda )  e^{\nu(\lambda )} \left(  \int d\lambda' {\cal P}_L( \lambda' )  e^{\nu(\lambda' )}     \right)^{K^L-1}  
\nonumber \\
\frac{\partial^2   {\cal G}_L [ \nu(.)  ] }{\partial^2 \nu(\lambda ) }   && 
=  \sum_{ {\cal Q}_L(.) } P_L [ {\cal Q}_L(.)  ] \ \left( K^L  {\cal Q}_{L} (\lambda) \right)^2
 \ e^{ \displaystyle  K^L \int d\lambda'  \nu(\lambda')  {\cal Q}_{L} (\lambda') }   
= K^L  {\cal P}_L( \lambda )  e^{\nu(\lambda )} \left(  \int d\lambda' {\cal P}_L( \lambda' )  e^{\nu(\lambda' )}    
 \right)^{K^L-1}  
\nonumber \\
&& +  K^L ( K^L-1) \left(  {\cal P}_L( \lambda )  e^{\nu(\lambda )} \right)^2 \left(  \int d\lambda' {\cal P}_L( \lambda' )  e^{\nu(\lambda' )}    
 \right)^{K^L-2}  
\label{geneq1treederi}
\end{eqnarray}
gives the integer moments of the number ${\cal N}_{L} (\lambda)= K^L {\cal Q}_{L} (\lambda)$ of branches with some Lyapunov exponent $\lambda$ (Eq \ref{nempib})
by taking $\nu(.)=0$. The first moment 
\begin{eqnarray}
\overline{ {\cal N}_{L} (\lambda) } = \sum_{ {\cal Q}_L(.) } P_L [ {\cal Q}_L(.)  ] \ K^L  {\cal Q}_{L} (\lambda)
= \frac{\partial   {\cal G}_L [ \nu(.)  ] }{\partial \nu(\lambda ) } \bigg\vert_{\nu(.)=0} = K^L  {\cal P}_L( \lambda )
 ={\cal N}^{typ}_{L} (\lambda)
\label{nempibav}
\end{eqnarray}
coincides with the typical value ${\cal N}^{typ}_{L} (\lambda) $ of Eq. \ref{nempibtyp}.
The second moment
\begin{eqnarray}
\overline{ \left( {\cal N}_{L} (\lambda) \right)^2 } = \sum_{ {\cal Q}_L(.) } P_L [ {\cal Q}_L(.)  ] \ \left( K^L  {\cal Q}_{L} (\lambda) \right)^2
= \frac{\partial^2   {\cal G}_L [ \nu(.)  ] }{\partial^2 \nu(\lambda ) } \bigg\vert_{\nu(.)=0} = 
K^L  {\cal P}_L( \lambda )  
 +  K^L ( K^L-1) \left(  {\cal P}_L( \lambda )   \right)^2 
\label{nempibav2}
\end{eqnarray}
can be rewritten in terms of the typical value ${\cal N}^{typ}_{L} (\lambda) $ of Eq. \ref{nempibtyp} as
\begin{eqnarray}
\overline{ \left( {\cal N}_{L} (\lambda) \right)^2 } 
\simeq  {\cal N}^{typ}_{L} (\lambda) + \left({\cal N}^{typ}_{L} (\lambda) \right)^2
\label{nempibav2typ}
\end{eqnarray}
and will thus change of behavior at the values $\lambda^{\pm}$ introduced in Eq. \ref{subextensive}.
In the region $\lambda^- < \lambda < \lambda^+ $ 
where ${\cal N}^{typ}_{L} (\lambda)  $ is exponentially large, the second term dominates
over the first term that corresponds to a small fluctuation.
In the other regions where ${\cal N}^{typ}_{L} (\lambda)  $ is exponentially small, the first term dominates
and actually represents the very small probability to have a single rare event
\begin{eqnarray}
\overline{ \left( {\cal N}_{L} (\lambda) \right)^2 } 
&& \simeq  \left({\cal N}^{typ}_{L} (\lambda) \right)^2 \ \ \ {\rm for } \ \ \lambda^- < \lambda < \lambda^+
\nonumber \\
&& \simeq  {\cal N}^{typ}_{L} (\lambda)  \ \ \ {\rm for } \ \ \lambda < \lambda^- \ \  {\rm and } \ \ \lambda^+< \lambda
\label{nempibav2typinside}
\end{eqnarray}
This result can be generalized to arbitrary moments, as described in the context of the Random Energy Model \cite{rem}.

\subsection { Statistical properties of the empirical sums ${\cal S}_{L}(k) $ of Eq. \ref{calmk}  }

The disorder-averaged value of the empirical sum ${\cal S}_{L}(k) $ of Eq. \ref{calmkdef} reads
\begin{eqnarray}
\overline { {\cal S}_{L}(k) } =  \sum_{b=1}^{K^L}  \overline { \vert \tau_L [v_b(. )] \vert^k } 
= K^L \overline { \vert \tau_L [v(. )] \vert^k } 
\label{calmkdefav}
\end{eqnarray}
where the moments $\overline { \vert \tau_L [v(. )] \vert^k } $
of non-integer order $k$ for the product of random variables have been already discussed 
in Eq. \ref{momentsk}
\begin{eqnarray}
\overline{  \left\vert \tau_L  [ v(. ) ] \right\vert^k }  && 
= \int d\lambda e^{k L \lambda} {\cal P}_L( \lambda ) = \int d\lambda e^{ L \left[k \lambda -  I(\lambda) \right] } 
=  e^{L\phi(k)  }  
\label{momentskt}
\end{eqnarray}
in terms of the 
scaled cumulant generating function $\phi(k)  $ 
\begin{eqnarray}
\phi(k)  = \ln \left[ \overline{ \vert t_{v} \vert^k }   \right]
\label{phikps}
\end{eqnarray}
that corresponds to the Legendre transform of the rate function $I(\lambda)$.

On the other hand, Eq. \ref{calmk} yields that the sum ${\cal S}_{L}(k)  $ in a typical sample
can be computed from the empirical histogram in a typical sample (Eq. \ref{nempibtypsample})
\begin{eqnarray}
{\cal S}^{TypicalSample}_{L}(k) =
\int d\lambda e^{ k L \lambda} {\cal N}^{TypicalSample}_L(\lambda) 
= K^L  \int d\lambda e^{  L \left[  k \lambda   - I(\lambda) \right] }  \theta( \lambda^- \leq \lambda \leq \lambda^+  )
\label{sktypicalsample}
\end{eqnarray}
So the only difference with the averaged value (Eqs \ref{calmkdefav} and \ref{momentskt})
\begin{eqnarray}
\overline { {\cal S}_{L}(k) } = K^L \int d\lambda e^{ L \left[k \lambda -  I(\lambda) \right] } 
\label{calmkdefavres}
\end{eqnarray}
lies in the boundaries $ \lambda^- \leq \lambda \leq \lambda^+ $ for the integration over the Lyapunov exponent
that appear for the value in a typical sample (Eq \ref{sktypicalsample}) but that are absent in the averaged value of Eq. \ref{calmkdefavres}.
As a consequence, one needs to discuss 
the position of the saddle-point value $\lambda_k$ 
that governs the integral governing the averaged value of Eq. \ref{calmkdefavres}
\begin{eqnarray}
k =  I'(\lambda_k) 
\label{lambdak}
\end{eqnarray}
with respect to the two boundaries $\lambda^{\pm}$ of the integral 
governing the typical-sample value
of Eq. \ref{sktypicalsample}.
It is thus useful to introduce the two values $k^{\pm}$ satisfying $\lambda_{k^{\pm}} = \lambda^{\pm} $
i.e.
\begin{eqnarray}
k^{\pm} =  I'(\lambda^{\pm}) 
\label{lambdakpm}
\end{eqnarray}
 and to distinguish the three following cases :

(a) In the region $ k^- < k < k^+  $,
 the saddle-point value $\lambda_k$ of Eq. \ref{lambdak} is in the interval
\begin{eqnarray}
 \lambda^- < \lambda_k < \lambda^+  
\label{lkinside}
\end{eqnarray}
The typical-sample value ${\cal S}^{TypicalSample}_{L}(k) $ of Eq. \ref{sktypicalsample}
 has then the same exponential behavior in $L$ as the averaged value $\overline { {\cal S}_{L}(k) } $
involving the Legendre transform $\phi(k)$ (Eqs \ref{momentskt} and \ref{phikp} ) of $I(\lambda)$
\begin{eqnarray}
{\cal S}^{TypicalSample}_{L}(k) \oppropto_{L \to +\infty}  K^L e^{L \phi(k)} = e^{ L \left[ \ln K + \phi(k)  \right] } 
\label{sktypicalsampleinside}
\end{eqnarray}

(b) In the region $k> k^+   $,
 the saddle-point value $\lambda_k$ of Eq. \ref{lambdak} is bigger than $\lambda^+$
\begin{eqnarray}
\lambda_k > \lambda^+ 
\label{lkright}
\end{eqnarray}
The typical-sample value ${\cal S}^{TypicalSample}_{L}(k) $ of Eq. \ref{sktypicalsample}
is then governed  by the saddle-point evaluation frozen at the boundary 
$\lambda_+$ satisfying $I(\lambda_+)=\ln K$ (Eq \ref{subextensive})
\begin{eqnarray}
{\cal S}^{TypicalSample}_{L}(k) \oppropto_{L \to +\infty} 
 K^L   e^{  L \left[  k \lambda^+   - I(\lambda^+) \right] }  =  e^{  L  k \lambda^+   }
\label{sktypicalsampleright}
\end{eqnarray}

(c) In the region $k< k^-   $,
  the saddle-point value $\lambda_k$ of Eq. \ref{lambdak} is smaller than $\lambda^-$
\begin{eqnarray}
\lambda_k < \lambda^- 
\label{lkleft}
\end{eqnarray}
 The typical-sample value ${\cal S}^{TypicalSample}_{L}(k) $ of Eq. \ref{sktypicalsample}
is then governed by the saddle-point evaluation frozen at the boundary 
$\lambda_-$ satisfying $I(\lambda_-)=\ln K$ (Eq \ref{subextensive})
\begin{eqnarray}
{\cal S}^{TypicalSample}_{L}(k) \oppropto_{L \to +\infty} 
 K^L   e^{  L \left[  k \lambda^-   - I(\lambda^-) \right] }  =  e^{  L  k \lambda^-   }
\label{sktypicalsampleleft}
\end{eqnarray}

\subsection { Sample-to-sample fluctuations in the frozen phase $k>k^+$ }

In the frozen phase $k>k^+$, the sample-dependent version of Eq. \ref{sktypicalsampleright}
is that the sum ${\cal S}_{L}(k)  $ in a given sample will be actually governed by the biggest
Lyapunov exponents available among the $K^L$ branches.
It is thus convenient to relabel in each sample the Lyaponov exponents according to their magnitudes
\begin{eqnarray}
\lambda_1 \geq \lambda_2 \geq \lambda_3 ... \geq \lambda_{K^L}
\label{lambdarelabel}
\end{eqnarray}
and to analyze the statistics of the first biggest terms in the sum of Eq. \ref{calmk}
\begin{eqnarray}
{\cal S}_{L}(k) = \sum_{b=1}^{K^L} e^{ k L \lambda_b} =  e^{ k L \lambda_1} +  e^{ k L \lambda_2} +  e^{ k L \lambda_3}+...
\label{calmkorder}
\end{eqnarray}
and in particular the first one that involves the maximal Lyapunov exponent $\lambda_1$
\begin{eqnarray}
{\cal S}^{first}_{L}(k) \equiv   e^{  L  k \lambda_1  }
\label{slkfirst}
\end{eqnarray}

\subsubsection { Probability distribution of the maximal Lyapunov exponent $\lambda_{1} $ in each sample  }

The maximal Lyapunov exponent $\lambda_{1} $ is typically of order $\lambda^+$, but here we wish to analyze its 
 probability distribution $R(\lambda_1) $ over the samples.
The corresponding cumulative distribution reads in terms of ${\cal P}_L(\lambda)$ of Eq. \ref{largedevlyapunovb}
\begin{eqnarray}
\int^{\lambda} d\lambda_1 R(\lambda_1) && = 
\left[ 1- \int_{\lambda}^{+\infty} d\lambda'  {\cal P}_L(\lambda' ) \right]^{K^L}
\opsimeq_{L \to + \infty}  e^{- K^L \int_{\lambda}^{+\infty} d\lambda'  e^{-L I(\lambda') } }
\opsimeq_{L \to + \infty}  e^{-   e^{ L( \ln K - I(\lambda) ) } }
\label{cumulqmax}
\end{eqnarray}

The change of variables 
\begin{eqnarray}
\lambda = \lambda^+ + \frac{u}{L k^+}
\label{defu}
\end{eqnarray}
centered around the value $\lambda^+ $ where $ I(\lambda^+)=\ln K $ and $ I(\lambda^+)=k^+ $ (Eq. \ref{lambdakpm})
leads to the Taylor expansion of the rate function 
\begin{eqnarray}
I(\lambda) = I(\lambda^+)  + \frac{u}{L k^+ } I'(\lambda^+)+ O \left( \frac{1}{L^2} \right) = \ln K 
+ \frac{u}{L } + O \left( \frac{1}{L^2} \right)
\label{bldef}
\end{eqnarray}
 Plugging this expansion into Eq \ref{cumulqmax} 
\begin{eqnarray}
\int^{\lambda^+ + \frac{u}{L k^+}} d\lambda_1 R(\lambda_1) 
\opsimeq_{L \to + \infty}  e^{- e^{ -  u  } } \equiv \int^u du'G(u')
\label{cumulqmaxu}
\end{eqnarray}
yields the convergence towards the Gumbel distribution 
(well-known as one of the three universality classes for the extreme-value statistics 
of independent random variables \cite{gumbel,galambos})
\begin{eqnarray}
G(u) = e^{-u}   e^{- e^{ -  u  } } 
\label{gumbel}
\end{eqnarray}
for the $O(1)$ random variable $u$ introduced in Eq. \ref{defu}.

\subsubsection { Probability distribution of ${\cal S}^{first}_{L}(k) =   e^{  L  k \lambda_1  }$ over the samples  }

Eq \ref{slkfirst} yields that its logarithm reads with the change of variables of Eq. \ref{defu}
\begin{eqnarray}
\ln {\cal S}^{first}_{L}(k) =   L  k \lambda_1   = L  k \lambda^+ +    \frac{k }{ k^+} u
\label{lnslkfrozen}
\end{eqnarray}
where $u$ is distributed with the Gumbel distribution of Eq. \ref{gumbel}.
This means that the probability distribution of $ \left( \ln {\cal S}^{first}_{L}(k) \right)$ propagates as a traveling wave
as $L$ grows : the first term $L  k \lambda^+ $ corresponds to a motion with the non-random velocity $( k \lambda^+)$
with respect to $L$, while the second term $\frac{k }{ k^+} u $ is random and independent of $L$,
i.e. its probability distribution corresponds to the fixed shape of the traveling wave.
This notion of traveling wave has been stressed here because it plays a major role 
in the analysis of random models defined on Cayley trees,
as first discovered with the exact solution of the Directed Polymer on the Cayley tree \cite{Der_Spo}.

Eq. \ref{lnslkfrozen} translates into
\begin{eqnarray}
 {\cal S}_{L}^{first}(k) = e^{L  k \lambda^+}  e^{    \frac{k }{ k^+} u } 
\equiv  \left( {\cal S}^{TypicalSample}_{L}(k) \right) X
\label{nslkfrozenu}
\end{eqnarray}
where  $ {\cal S}^{TypicalSample}_{L}(k)=e^{L  k \lambda^+} $ is the value in a typical sample introduced in
\ref{sktypicalsampleright}, while 
\begin{eqnarray}
X \equiv e^{    \frac{k }{ k^+} u } 
\label{defsigmak}
\end{eqnarray}
is an $O(1)$ positive random variable, whose distribution reads in terms of the 
Gumbel distribution $G(u)$ of Eq. \ref{gumbel} 
\begin{eqnarray}
L_k( X ) = \frac{k^+}{k X }  G \left( \frac{k^+}{k } \ln X \right) 
=  \frac{ \mu_k }{ X^{1+\mu_k} } e^{ - \frac{ 1 }{ X^{\mu_k} } }
\label{levymu}
\end{eqnarray}
where the exponent
\begin{eqnarray}
\mu_k \equiv \frac{ k^+ }{k} 
\label{defmuk}
\end{eqnarray}
 governs the power-law decay of Eq. \ref{levymu} for large $X$
\begin{eqnarray}
L_k( X ) \opsimeq_{X \to + \infty}   \frac{ \mu_k }{ X^{1+\mu_k} } 
\label{levymutail}
\end{eqnarray}
The exponent $\mu_k $ decays continuously in the frozen phase $k \geq k^+$ from the value $\mu_{k=k^+} =1 $
to vanishing values $\mu_{(k \to +\infty)} \to 0$.
Since it remains smaller than one in the whole frozen phase $k \geq k^+ $
\begin{eqnarray}
\mu_{(k \geq k^+)} = \frac{ k^+ }{k} \leq 1
\label{musmaller}
\end{eqnarray}
the averaged value of the variable $X$ is infinite
\begin{eqnarray}
\int dX  X L_k( X ) = +\infty
\label{levyavinfty}
\end{eqnarray}
i.e. the averaged value $\overline{  {\cal S}_{L}^{first}(k) }$ in Eq. \ref{nslkfrozenu}
has a different exponential behavior in $L$ than the typical value ${\cal S}^{TypicalSample}_{L}(k) $,
in consistency with the discussion around Eq. \ref{calmkdefavres}.


\subsection { Application to the Directed Polymer and the Random Energy Model }

With respect to the generic notations of section \ref{sec_star},
the Random Energy Model \cite{rem} corresponds to $K=2$ 
and to the case where $\lambda$ is an energy distributed with a Gaussian distribution of Eq. \ref{gauss}
so that the rate function $I(\lambda)$ and the scaled cumulant generating function $\phi(k)  $ are quadratic
\begin{eqnarray}
I^{Gauss}(\lambda) && =  \frac{\lambda^2}{2   \sigma^2} 
\nonumber \\
\phi^{Gauss}(k) && =  \frac{ k^2  \sigma^2}{2} 
\label{largedevgausssansbeta}
\end{eqnarray}
The empirical number $ {\cal N}_{L} (\lambda)$ of Eq. \ref{nempib} corresponds to the number of accessible states
in the microcanonical ensemble where the energy density $\lambda$ is fixed,
and its value $ {\cal N}_{L} (\lambda) $ in a typical sample (Eq. \ref{nempibtypsample})
yields that the function in the exponential corresponds to the entropy as a function of the energy density $\lambda$
in the microcanonical ensemble \cite{rem}
\begin{eqnarray}
S^{Microcanonical}(\lambda) = \left[ \ln K -  I^{Gauss}(\lambda) \right]  \theta( \ln K -  I^{Gauss}(\lambda) \geq 0  )
= \left[ \ln 2 -  \frac{\lambda^2}{2   \sigma^2}  \right]  \theta( \lambda^- \leq \lambda \leq \lambda^+  )
\label{smicrotypsample}
\end{eqnarray}
with the boundaries
\begin{eqnarray}
\lambda^{\pm} = \pm  \sigma \sqrt{ 2 \ln 2 }
\label{smicrolpm}
\end{eqnarray}
With the change of notation $k \to \beta$, the empirical sum ${\cal S}_{L}(k) $ of Eq. \ref{calmkdef} and \ref{calmk}
corresponds to the partition function $Z_{L}(\beta)  $ in the canonical ensemble at inverse temperature $\beta$
\begin{eqnarray}
Z_{L}(\beta) = \sum_{b=1}^{K^L} e^{ \beta L \lambda_b} = \int d\lambda e^{ k L \lambda} {\cal N}_L(\lambda) 
\label{zrem}
\end{eqnarray}
with its disordered-averaged value (Eqs \ref{calmkdefav} and \ref{momentskt})
\begin{eqnarray}
\overline { Z_{L}(\beta) } 
= K^L   e^{L\phi(\beta)  }  =  e^{L \left[ \ln 2 +\frac{ \beta^2  \sigma^2}{2} \right]  }
\label{zremav}
\end{eqnarray}
while its value in a typical sample (Eq \ref{sktypicalsample}) involves the microcanonical entropy of Eq. \ref{smicrotypsample}
\begin{eqnarray}
Z^{TypicalSample}_{L}(\beta) 
= K^L  \int d\lambda e^{  L \left[  \ln 2 -  \frac{\lambda^2}{2   \sigma^2}   \right] } 
\theta( \lambda^- \leq \lambda \leq \lambda^+  )
\label{ztypicalsample}
\end{eqnarray}

Since the inverse temperature $\beta$ is positive $\beta>0$ (instead of $k$ of arbitrary sign above),
the critical temperature $\beta_c$ of the freezing transition corresponds to the solution $k^+$ of Eq \ref{lambdakpm}
\begin{eqnarray}
\beta_c =  I'(\lambda^{+}) = \frac{\lambda^+}{   \sigma^2} =  \frac{  \sqrt{ 2 \ln 2 }}{   \sigma} 
\label{betacrem}
\end{eqnarray}
The two phases are \cite{rem}

(a) the high-temperature phase $\beta<\beta_c$
where the partition function in a typical sample (Eq \ref{sktypicalsample}) coincides with the averaged value of Eq. \ref{zremav}.

(b) the low-temperature frozen phase $\beta<\beta_c$
where the partition function in a typical sample is different from the averaged value of Eq. \ref{zremav}
because it is governed by the boundary $\lambda_+$ (Eq. \ref{sktypicalsampleright})
\begin{eqnarray}
Z^{TypicalSample}_{L}(\beta) \oppropto_{L \to +\infty}   e^{  L  \beta \lambda^+   } = e^{  L  \beta  \sigma \sqrt{ 2 \ln 2 }  } 
\label{zremtypfrozen}
\end{eqnarray}
In this frozen phase, the exponent of Eq. \ref{defmuk}
\begin{eqnarray}
\mu_{\beta} \equiv \frac{ \beta_c }{ \beta} = \frac{T}{T_c}
\label{mubeta}
\end{eqnarray}
of the heavy-tail distribution of Eq. \ref{levymutail}
allows to analyze further the statistics of overlaps
in terms of the weights of individual terms 
within in a L\'evy sum of random variables distributed with heavy tails \cite{Der_Tou,Der_Fly,Der_review}.


\subsection{ Application to Anderson Localization  }

The notations for the Anderson Localization model have been explained in the subsection \ref{exampleAnderson}
with the rate function $I(\lambda)$ and the scaled cumulant generating function $\phi(k)$
given by Eqs \ref{phikanderson}
and \ref{ianderson}
\begin{eqnarray}
I^{Anderson}(\lambda) && = \lambda - \ln \left( \frac{ V}{ W}\right) -1 -\ln \left[ \lambda - \ln \left( \frac{ V}{ W}\right) \right]
\nonumber \\
\phi^{Anderson}(k) && = k \ln \left( \frac{ V}{ W}\right) - \ln (1-k)
\label{iphianderson}
\end{eqnarray}
Here the analysis concerns the localized phase in the regime of small hopping $V$ where
the forward perturbation formula of Eq. \ref{psi0r} is valid,
so it will be possible to use this approach up to the critical hopping $V_c$ of the delocalization transition
only if the branching ratio $K$ is large $K \gg 1$.

The empirical number of Eq \ref{nempib} counts the number of leaves (among the $K^L$ branches)
where the wave-function $\vert \psi_b(L)\vert $ is of order $e^{L \lambda}$ with respect
to the finite wave-function at the center.
The empirical number in a typical sample (Eq. \ref{nempibtypsample}) reads
\begin{eqnarray}
{\cal N}^{TypicalSample}_{L} (\lambda)  \opsimeq_{ L \to + \infty} 
  e^{ L \left[\ln K -  I^{Anderson}(\lambda) \right] } \theta( \lambda^- \leq \lambda \leq \lambda^+  )
= e^{ L \left[ \ln \left(  \frac{ K V e}{ W}\right) 
- \lambda  +\ln \left[ \lambda - \ln \left( \frac{ V}{ W}\right) \right]  \right] } \theta( \lambda^- \leq \lambda \leq \lambda^+  )
\label{nempibtypsampleanderson}
\end{eqnarray}
where the boundaries $\lambda^{\pm}$ are given by Eq. \ref{subextensive}
\begin{eqnarray}
0= \ln K - I^{Anderson}(\lambda^{\pm} )= \ln \left(  \frac{ K V e}{ W}\right) 
- \lambda^{\pm}  +\ln \left[ \lambda^{\pm} - \ln \left( \frac{ V}{ W}\right) \right]
\label{subextensiveanderson}
\end{eqnarray}
For large $K\gg 1$, the upper boundary is given by
\begin{eqnarray}
\lambda^+ \simeq \ln \left(  \frac{ K V e}{ W}\right) 
 +\ln \left[ \ln \left(  \frac{ K V e}{ W}\right)    - \ln \left( \frac{ V}{ W}\right) \right]
= \ln \left[  \frac{ K V e}{ W} \ln (Ke) \right] 
\label{lambdaplusanderson}
\end{eqnarray}
The localized phase correspond to the region $\lambda_+<0$,
where the wave-function decays exponentially on all the $K^L$ branches,
while the delocalization transition occurs when $\lambda^+$ vanishes 
\begin{eqnarray}
  (\lambda^+)_{criti} =0
\label{critianderson}
\end{eqnarray}
so the critical hopping $V_c$ for the delocalization transition is given for large $K\gg 1$ by
\begin{eqnarray}
 V_c  \simeq \frac{W}{K  e \ln (Ke) } 
\label{vcanderson}
\end{eqnarray}
At this critical point $V=V_c$, the inverse participation ratios 
\begin{eqnarray}
Y_{L}(q) \equiv \sum_{b=1}^{K^L}     \vert \psi_b(L)  \vert^{2q}
\label{yqdef}
\end{eqnarray}
correspond to the empirical sums of Eq. \ref{calmkdef} with the change of notation $k=2q$,
so that their disordered-averaged values (Eqs \ref{calmkdefav} and \ref{momentskt}) read for $q<\frac{1}{2}$
\begin{eqnarray}
\overline { Y_{L}(q) } 
= K^L   e^{L\phi^{Anderson}(2q)  }  =  e^{ L \left[ \ln K + 2q \ln \left( \frac{ V_c}{ W}\right) - \ln (1-2q) \right] }
=  e^{ L \left[ (1-2q) \ln K - 2q   \ln (e \ln (Ke) ) - \ln (1-2q) \right] } \opsimeq_{L \to +\infty} (K^L)^{-\tau^{av}_q}
\label{yqavav}
\end{eqnarray}
where the exponents $\tau_q^{av}$ defined with respect to the number $K^L$ of sites read for large $K\gg 1$
\begin{eqnarray}
\tau^{av}_q \simeq  (2q-1) \theta \left( q< \frac{1}{2}\right)
\label{tauq}
\end{eqnarray}
Eq \ref{lambdakpm} yields that the boundary value $k^+=2q^+$
using Eq. \ref{lambdaplusanderson} and Eq. \ref{vcanderson}
\begin{eqnarray}
2 q^{+} =  I'(\lambda^{+}) = 1+
 \frac{1}{  \lambda^+ - \ln \left( \frac{ V_c}{ W}\right) }
= 1 - \frac{1}{  \ln \left[   K  e \ln (Ke) \right]   }
\label{qpm}
\end{eqnarray}
is close to unity for large $K$, so that the inverse participation ratios in a typical sample
\begin{eqnarray}
Y_{L}^{TypicalSample}(q) \opsimeq_{L \to +\infty} (K^L)^{-\tau^{typ}_q}
\label{yqdeftyp}
\end{eqnarray}
involve essentially the same exponents as the averaged values of Eq. \ref{yqavav}
\begin{eqnarray}
\tau^{typ}_q \simeq \tau^{av}_q \simeq  (2q-1) \theta \left( q< \frac{1}{2}\right)
\label{tauqtyp}
\end{eqnarray}
These exponents are known as the 'Strong Multifractality spectrum' in the field of Anderson transitions \cite{mirlinrevue},
where they appears either in the limit of infinite dimensionality $d \to +\infty$
or in related long-ranged power-law hoppings in one-dimension 
 \cite{levitov1,levitov2,levitov3,levitov4,mirlin_evers,fyodorov,fyodorovrigorous,oleg1,oleg2,oleg3,oleg4,olivier_per,olivier_strong,olivier_conjecture,us_strongmultif},
or more recently in toy models of Many-Body-Localization \cite{c_mblperturb,c_mblstrongmultif}.
Although the freezing transitions at the values $q^{\pm}$ is not very important for this 
'Strong Multifractality spectrum', they have been much discussed in the general theory of multifractality
at Anderson transitions in finite dimension $d$ \cite{janssenrevue,mirlinrevue,mirlin_evers}

\subsection { Application to the Quantum Ising Model }

As a third and final example, let us mention the case of the random transverse field spin-glass model on the Cayley tree
that has been studied recently via real-space renormalization 
and where the large deviations properties of the one-dimensional model play a major role \cite{mblcayley}.
Here the difference with the two previous examples of the Random Energy Model and of Anderson Localization
is that the one-dimensional model has already its phase transition between the spin-glass phase and the paramagnetic phase, where the exact critical properties have been obtained by the Strong Disorder renormalization approach \cite{sdrgreview,sdrgminireview}. As a consequence, one obtains three phases that can be explained as follows
in the star-geometry (ii) of Eq. \ref{nlstar}, where one considers that the center $O$
is linked to $K^L$ independent chains of length $L$ \cite{mblcayley}, i.e. each branch $b=1,..,K^L$
is characterized by the Lyapunov exponent (Eq \ref{crq})
 \begin{eqnarray}
\lambda_b \equiv \frac{1}{L} \sum_{x=1}^L \ln \left\vert \frac{J_b(x)}{h_b(x)} \right\vert
\label{lambdabqi}
\end{eqnarray}

(a) the star is in its paramagnetic phase if all the $K^L$ chains are in their paramagnetic state $\lambda_b<0$,
i.e. the boundary value $\lambda_+$ of Eq. \ref{subextensive} should be negative $\lambda_+<0$.

(b) the star is in its spin-glass phase with an extensive spin-glass order
if an extensive number of the $K^L$ chains are in their spin-glass phase $\lambda_b>0$
i.e. the typical Lyapunov exponent should be positive $\lambda_{typ}>0$.

(c) in between, i.e. in the region $\lambda_{typ}<0<\lambda_+$,
 the star is a spin-glass phase with an sub-extensive spin-glass order,
because only the subextensive number of chains 
are in their spin-glass phase $\lambda_b>0$, while an extensive number of chains
are in their paramagnetic state $\lambda_b<0$.


\section{ Conclusion}

\label{sec_conclusion}

In this pedagogical introduction, we have explained 
why the general theory of large deviations is the natural language to analyze the properties
of disordered systems in order to offer
 a unified perspective on the typical events and on the rare events that occur on various scales.
 We have first focused on one-dimensional random models
in order to emphasize the various levels of description.
We have first recalled how the Level 1 allows to analyze the properties of
observables given by products of random variables 
that occur in many classical or quantum models.
We have then described how a finer analysis in terms of 
the whole hierarchy of empirical histograms 
allows to classify the set of disorder configurations
into subsets that have the same empirical properties up to a certain order.
We have then turned our attention to random models defined on Cayley trees,
in order to analyze their properties in terms of the large deviations of branches.
We have taken as examples various emblematic classical and quantum disordered systems
 in order to highlight the common underlying mechanisms from the point of view of large deviations. 

The large deviation analysis of disordered systems in finite dimension $1<d<+\infty$ 
 clearly goes beyond the scope of the present introduction.
Although some notions can be directly applied, like the Sanov theorem for the empirical 1-point histogram,
or the multifractal analysis at Anderson transitions \cite{janssenrevue,mirlinrevue} 
or at phase transitions of random classical models \cite{Ludwig,Jac_Car,Cha_Ber,Cha_Ber_Sh,PCBI,BCrevue,Thi_Hil,symetriemultif},
one should be aware that qualitatively new phenomena may also occur.
For instance the large deviations properties that have been exactly computed \cite{derrida_leb,dean_maj,maj_verg}
for the Directed Polymer in dimension $d=2$ display an
 asymmetry between values bigger or smaller than the typical value, 
with two different scalings with respect to the length $L$ of the polymer  :
an 'anomalously good' ground state energy requires only $L$ anomalously good on-site energies along the polymer,
while an 'anomalously bad' ground state energy requires $L^2$ bad on-site energies in the two-dimensional sample. 
So this single example already shows that some properties of random systems in finite dimensions $d$
call for a much broader large deviation theory with two different scalings for values bigger or smaller than the typical value,
as discussed in more details in the recent preprint \cite{largedevAsym}.


\appendix

\section{ Alternative classification of disorder configurations in terms of empirical intervals }

\label{sec_interval}

In the text, we have described the classification of one-dimensional disorder configurations
in terms of the hierarchy of the empirical r-point histograms.
In this Appendix, we discuss an alternative classification in terms of the empirical intervals 
during which the disorder keeps a constant value, since this framework is 
more appropriate to analyze the Lifshitz and the Griffiths singularities as we now recall.

\subsection{ Observables corresponding to products of contributions from intervals of random lengths }

\label{sec_obsinterval}

After the product of random variables discussed in section \ref{sec_obsprod},
the next simpler case of Eq. \ref{defprod} concerns
the case where the disorder variable can take only two values
that will be labelled by $v=\pm$. It is then useful to replace the disorder 
configuration $[v(x )]_{x=1,2,..,L}$
by its decomposition into intervals during which the disorder keeps the same value.
For a model defined on a ring of $L$ sites
 (i.e. with periodic boundary conditions $L+x=x$) there will be an empirical even number $(2N)$
of intervals, where the $N$ odd intervals $(2i-1)$ of lengths $l_{2i-1}$ are associated to the value $v=-$,
while the $N$ even intervals $(2i)$ of lengths $l_{2i}$ are associated to the value $v=+$.
The lengths $l_i$ satisfy the sum rule
\begin{eqnarray}
L = \sum_{i=1}^{2N} l_i
\label{tsum}
\end{eqnarray}
When the disorder configuration $[v(x )]_{x=1,2,..,L}$
is replaced by the list $\left[ l_i \right]_{i=1,2,..,2N}$ of the lengths  of the intervals,
the trace of Eq. \ref{defprod} becomes
\begin{eqnarray}
\Theta_L \left( \left[ l_i \right]_{i=1,2,..,2N} \right) \equiv 
Tr\left[  T_{ + }^{l_{2N}}  T_{-}^{l_{2N-1}} ...  T^{l_3}_{-} T_{+}^{l_2} T^{l_1}_{-} 
\right]
\label{zinterval}
\end{eqnarray}

To analyze the Lifshitz and the Griffiths singularities mentioned in the Introduction,
various models have been studied in the regime where the value $v=-$
 corresponds to a very strong disorder value where the associated transfer matrix $T_-$
can be approximated by a projector on some state $\vert 0>$ with some eigenvalue $t_-$ \cite{luckbook}
\begin{eqnarray}
T_- \simeq  t_- \vert 0 > < 0 \vert 
\label{tmoinsproj}
\end{eqnarray}
Then Eq. \ref{zinterval} simplifies into the product of the contributions of the intervals
\begin{eqnarray}
\Theta_L \left( \left[ l_i \right]_{i=1,2,..,2N} \right) 
= \prod_{i=1}^N \left( \theta_+(l_{2i})  \ \theta_-(l_{2i-1}) \right)
\label{zprodinterval}
\end{eqnarray}
where the contribution of an interval $v=-$ of length $l$ 
is simply
\begin{eqnarray}
\theta_-(l) \equiv  (t_-)^l
\label{zmoinsl}
\end{eqnarray}
while the contribution of an interval $v=+$ of length $l$ 
corresponds to the the pure model $v=+$ with the boundary conditions
$\vert 0>$ fixed by the projector form of Eq. \ref{tmoinsproj}
\begin{eqnarray}
\theta_+(l) \equiv  < 0 \vert T_+^{l} \vert 0 > 
\label{zplusl}
\end{eqnarray}
Various examples concerning Anderson Localization models and classical spin chains 
are described in the book \cite{luckbook}, while an example concerning random DNA is analyzed in \cite{kafri}.

\subsection{ Empirical 1-interval observables with their constraints }

The observables of the form of Eq. \ref{zprodinterval}
 suggests that it is appropriate to analyze the disorder configurations
in terms the empirical 1-interval observables
\begin{eqnarray}
n_{+} (l) && \equiv \frac{1}{L} 
\sum_{i=1}^N  \delta_{l_{2i}, l}
\nonumber \\
n_{-} (l) && \equiv \frac{1}{L} 
\sum_{i=1}^N  \delta_{l_{2i-1}, l} 
\label{n1empi}
\end{eqnarray}

The summation over the length $l$ corresponds to the density $\frac{N}{L} $ of intervals $v=+$ or $v=-$
\begin{eqnarray}
\sum_{l} n_{+} \left( l \right) && = \frac{N}{L} = \sum_{l} n_{-} \left( l \right)
\label{n1empit}
\end{eqnarray}
while the total length $L$ of the disorder configurations fixes the normalization (Eq. \ref{tsum})
\begin{eqnarray}
1=  \sum_{l=1}^{+\infty} l \left[ n_{+} \left( l \right)  + n_{-} \left( l \right) \right]
\label{ltota}
\end{eqnarray}
It is thus useful to introduce the following notation to summarize these constraints
on the empirical 1-interval observables $n^{\pm}(.)$
\begin{eqnarray}
 c_1[ n_+(.); n_-(.)]  \equiv 
&& \delta \left(  \sum_{l} n_{+} \left( l \right)  -\sum_{l} n_{-} \left( l \right)  \right)
\delta \left( 1 - \sum_{l} l \left[ n_{+} \left( l \right)  + n_{-} \left( l \right) \right] \right)
\label{c1inter}
\end{eqnarray}
where again the notation $\delta(X)$ is introduced for better readability of the arguments $X$
but actually represents the Kronecker symbol $\delta_{0,X}$.

\subsection{ Typical values of the empirical 1-interval observables }

\label{subsec_Lypical}

Since the probability of a disorder configuration is given by Eq. \ref{ptrajindep} with $p_+ + p_-=1$,
the probability distributions of the lengths $l$ of the intervals $v=\pm$ 
are given by the geometrical distributions
\begin{eqnarray}
p^{geo}_{\pm}(l) = (1-p_{\pm} ) (p_{\pm})^{l-1}
\label{ppmgeom}
\end{eqnarray}
with the normalization
\begin{eqnarray}
\sum_{l=1}^{+\infty}  p^{geo}_{\pm}(l) =1
\label{pal}
\end{eqnarray}
and the averaged lengths
\begin{eqnarray}
\sum_{l=1}^{+\infty}  l p^{geo}_{\pm}(l) = \frac{1}{1-p_{\pm} } =  \frac{1}{p_{\mp} } 
\label{tauav}
\end{eqnarray}
As a consequence, the typical density $\frac{N^{typ}}{L}$ of the intervals reads
\begin{eqnarray}
\frac{N^{typ}}{L} = \frac{1}{ \displaystyle \sum_{l=1}^{+\infty}  l \left[ p^{geo}_{+}(l) + p^{geo}_{-}(l)  \right]  } = p_+ p_-
\label{ntyp}
\end{eqnarray}
and the typical values of the empirical 1-interval observables are
\begin{eqnarray}
n_{+}^{typ} (l) && = \frac{N^{typ}}{L} p^{geo}_{+}(l) =  p_-^2  p_{+}^{l}
\nonumber \\
n_{-}^{typ} (l) && =\frac{N^{typ}}{L} p^{geo}_{-}(l) = p_+^2  p_{-}^{l}
\label{n1empityp}
\end{eqnarray}

\subsection{ Large deviations of empirical 1-interval observables }

In order to analyze the large deviations of empirical 1-interval observables,
one needs to introduce a generalized semi-Markovian model for the disorder,
where the lengths $l_i$ of the intervals are drawn with some general distributions $p_{\pm}(l)$
(instead of the geometric distributions of Eq. \ref{ppmgeom}).
The probability of some configuration of the intervals then reads
(up to boundary terms that can be neglected for $L \to +\infty$
\begin{eqnarray}
{\mathbb P}^{SemiMarkov}_L \left[ (l_{2N} ...l_1)\right]  
&& \simeq 
   \delta \left( L -  \sum_{i=1}^{2N} l_i \right) \prod_{i=1}^N p_+(l_{2i}) p_-(l_{2i-1}) 
= \delta \left( L -  \sum_{i=1}^{2N} l_i \right) e^{ \sum_{i=1}^N \left[ \ln (p_+(l_{2i}) ) + \ln ( p_-(l_{2i-1}) ) \right] } 
\nonumber \\
&& \simeq  c_1[ n_+(.); n_-(.)]     e^{-L a [ n_{+}(.); n_{-}(.) ] }
\label{paction}
\end{eqnarray}
where the action in the exponential is a function of the empirical 1-interval observables introduced in Eq. \ref{n1empi}
\begin{eqnarray}
a [ n_{+}(.); n_{-}(.) ] = -   \sum_{l}
\left[  n_{+}( l)  \ln \left( p_{+}(  l) \right)
+ n_{-}( l)  \ln \left( p_{-}(  l) \right)
\right]
\label{action}
\end{eqnarray}
while $c_1[ n_+(.); n_-(.)] $ has been introduced in Eq. \ref{c1inter} to summarize the constraints.
In this semi-Markovian model, all the disorder configurations 
that have the same empirical 1-interval observables $n^{\pm}(.)$
have the same probability.
As a consequence, the probability $P_L[n_{+}(.); n_{-}(.)  ]$ to see these empirical observables
is given by
\begin{eqnarray}
P_L[ n_{+}(.); n_{-}(.)  ]
 =  c_1[ n_+(.); n_-(.)]  \omega_L[  n_{+}(.); n_{-}(.)  ]  e^{ - L a [ n_{+}(.); n_{-}(.) ]     }
\label{probaempi}
\end{eqnarray}
where $\omega_L[  n_{+}(.); n_{-}(.)  ]  $ counts the 
number of disorder configurations 
 that correspond to these empirical observables, while the normalization reads
\begin{eqnarray}
 1 && =
\left(  \prod_{l} \int dn_+(l) \int dn_-(l)  \right)  P_L[ n_{+}(.); n_{-}(.)  ]
\label{normamicro}
\end{eqnarray}

When the empirical 1-interval observables take their typical values for this semi-Markovian generalized 
model (adapted from Eqs \ref{ntyp} and \ref{n1empityp} )
\begin{eqnarray}
n_{\pm}^{typ} (l) && =\frac{ p_{\pm}(l) }{ \displaystyle \sum_{l'=1}^{+\infty}  l' \left[ p_{+}(l') + p_{-}(l')  \right]  }
\label{n1empitypsemi}
\end{eqnarray}
 the probability 
$P_L[ n^{typ}_{+}(.); n^{typ}_{-}(.)  ] $ should remain finite as $L \to +\infty$.
So the factor $\omega_L[ n^{typ}_{+}(.); n^{typ}_{-}(.)]  $ should compensate exactly the exponential factor 
of Eq. \ref{probaempi}, i.e. it should display 
 the exponential growth
\begin{eqnarray}
\omega_L[ n^{typ}_{+}(.); n^{typ}_{-}(.) ] &&  \opsimeq_{L \to +\infty} 
e^{ + L  a [ n^{typ}_{+}(.); n^{typ}_{-}(.) ]   }
= e^{ - \displaystyle L  
  \sum_{l}
\left[  n^{typ}_{+}( l)  \ln \left( p_{+}(  l) \right)
+ n^{typ}_{-}( l)  \ln \left( p_{-}(  l) \right)
\right]
   }
\label{microexptyp}
\end{eqnarray}

When the empirical observables $[n_{+}(.); n_{-}(.)] $ are different from their typical values
$[ n^{typ}_{+}(.); n^{typ}_{-}(.) ]$,
we may consider a modified semi-Markovian model with modified probability distributions ${\tilde p}_{\pm}(l)$
for the lengths of the intervals 
 that would make the empirical  observables $[n_{+}(.); n_{-}(.)] $
typical for this modified model. 
Equations \ref{n1empitypsemi} 
yield that the modified probability distributions ${\tilde p}_{\pm}(l)$ should be chosen as
\begin{eqnarray}
{\tilde p}_{\pm}( l)  && = \frac{n_{\pm} (l) }{ \sum_{l'} n_{\pm} (l')}  
\label{tildepinter}
\end{eqnarray}
where the two denominators coincide as a consequence of the constraints of Eq. \ref{c1inter}

Then Eq. \ref{microexptyp} translates for this modified model into
\begin{eqnarray}
\omega_L[ n_{+}(.); n_{-}(.) ] &&  \opsimeq_{L \to +\infty} 
 e^{ - \displaystyle L    \sum_{l}
\left[  n_{+}( l)  \ln \left( {\tilde p}_{+}(  l) \right)
+ n_{-}( l)  \ln \left( {\tilde p}_{-}(  l) \right) \right]
   }
\nonumber \\
&& =
 e^{ - \displaystyle L    \sum_{l}
\left[  n_{+}( l)  \ln \left( \frac{n_{+} (l) }{ \sum_{l'} n_{+} (l')}   \right)
+ n_{-}( l)  \ln \left( \frac{n_{-} (l) }{ \sum_{l'} n_{-} (l')}   \right)  \right]
   }
\label{microexptypmod}
\end{eqnarray}

Plugging this result into Eq. \ref{probaempi} yields the large deviation form
\begin{eqnarray}
P_L[ n_{+}(.); n_{-}(.)  ]
 \opsimeq_{L \to +\infty}   c_1[ n_+(.); n_-(.)]  
 e^{ -  L    J[ n_{+}(.); n_{-}(.) ]
   }
\label{probaempild}
\end{eqnarray}
with the rate function
\begin{eqnarray}
 J[ n_{+}(.); n_{-}(.) ] =  
  \sum_{l}
\left[  n_{+}( l)  \ln \left( \frac{n_{+} (l) }{ p_{+}(  l) \sum_{l'} n_{+} (l')}   \right)
+ n_{-}( l)  \ln \left( \frac{n_{-} (l) }{ p_{-}(  l)\sum_{l'} n_{-} (l')}   \right)  \right] 
\label{rateinterempi}
\end{eqnarray}
Related studies on large deviations properties of various semi-Markov processes in continuous time
can be found in \cite{fortelle_thesis,gaspard,maes_semi,zambotti,faggionato}.

Here we wish to return to the initial disorder model corresponding to the geometric distributions $p^{geo}_{\pm}(l) $ of Eq. \ref{ppmgeom}, where the result of Eq. \ref{rateinterempi}, concerning the generalized semi-Markov model
of disorder configurations with arbitrary distributions $p_{\pm}(l)$ for the lengths of the intervals,
becomes
\begin{eqnarray}
 J^{geo}[ n_{+}(.); n_{-}(.) ] =  
  \sum_{l}
\left[  n_{+}( l)  \ln \left( \frac{n_{+} (l) }{ p^{geo}_{+}(  l) \sum_{l'} n_{+} (l')}   \right)
+ n_{-}( l)  \ln \left( \frac{n_{-} (l) }{ p^{geo}_{-}(  l)\sum_{l'} n_{-} (l')}   \right)  \right] 
\label{rateinterempigeo}
\end{eqnarray}

\subsection{ Large deviations for observables given by the product of the intervals contributions   }

The modulus of Eq. \ref{zprodinterval} 
can be rewritten in terms of the empirical 1-interval observables of Eq. \ref{n1empi}
as
\begin{eqnarray}
\left\vert \Theta_L \left( \left[ l_i \right]_{i=1,2,..,2N} \right) \right\vert
= e^{ \displaystyle \sum_{i=1}^N \left[ \ln \left\vert \theta_+(l_{2i}) \right\vert +
\ln \left\vert \theta_-(l_{2i-1}) \right\vert
 \right]  }
= e^{  \displaystyle L \sum_{l} \left[ n_+(l) \ln \left\vert \theta_+(l) \right\vert 
+  n_-(l)  \ln \left\vert \theta_-(l) \right\vert 
\right]  }
\label{zprodintervalempi}
\end{eqnarray}
So the corresponding finite-size Lyapunov exponent of Eq. \ref{lyapunov}
is a linear function of the empirical 1-interval observables
\begin{eqnarray}
\lambda [ n_{+}(.); n_{-}(.) ]  = \frac{ \ln  \vert  \Theta_L  \vert  }{ L} 
= \sum_{l} \left[ n_+(l) \ln \left\vert \theta_+(l) \right\vert 
+  n_-(l)  \ln \left\vert \theta_-(l) \right\vert \right]
\label{lyapunovinter}
\end{eqnarray}

Its typical value can be obtained from the typical values of the empirical 1-interval 
observables of Eq. \ref{n1empityp}
\begin{eqnarray}
\lambda^{typ} && 
= \sum_{l} \left[ n_+^{typ}(l) \ln \left\vert \theta_+(l) \right\vert 
+  n_-^{typ}(l)  \ln \left\vert \theta_-(l) \right\vert \right]
=  \sum_{l} \left[ p_-^2  p_{+}^{l} \ln \left\vert \theta_+(l) \right\vert 
+  p_+^2  p_{-}^{l}  \ln \left\vert \theta_-(l) \right\vert \right]
\label{lyatyp}
\end{eqnarray}

The moments of non-integer order $k$ of Eq \ref{zprodintervalempi}
read in terms of the probability $P_L[ n_{+}(.); n_{-}(.)  ] $ of Eq \ref{probaempild}
\begin{eqnarray}
\overline{ \left\vert \Theta_L  \right\vert^k } && = 
\left(  \prod_{l} \int dn_+(l) \int dn_-(l)  \right) P_L[ n_{+}(.); n_{-}(.)  ] e^{L k \lambda [ n_{+}(.); n_{-}(.) ] } 
\nonumber \\
&&  \opsimeq_{L \to +\infty} 
\left(  \prod_{l} \int dn_+(l) \int dn_-(l)  \right)
  c_1[ n_+(.); n_-(.)]  
 e^{   L \left[   k \lambda [ n_{+}(.); n_{-}(.) ] -   J^{geo}[ n_{+}(.); n_{-}(.) ] \right]   }
\label{thetaqav}
\end{eqnarray}
One thus needs to optimize the function 
$\left[   k \lambda [ n_{+}(.); n_{-}(.) ] -   J^{geo}[ n_{+}(.); n_{-}(.) ] \right] $ in the exponential 
in the presence of the constraints $c_1[ n_+(.); n_-(.)]   $ of Eq. \ref{c1inter}
that can be taken into account via Lagrange multipliers.
It is technically more convenient to introduce the empirical density of intervals $\pm$
that appear in the constraints $c_1[ n_+(.); n_-(.)]   $ and in the rate function $ J^{geo}[ n_{+}(.); n_{-}(.) ]  $
of Eq. \ref{rateinterempigeo}
\begin{eqnarray}
n \equiv  \sum_{l} n_{+} \left( l \right) = \sum_{l} n_{-} \left( l \right)
\label{ndensity}
\end{eqnarray}
via another constraint.

So we will consider the functional
\begin{eqnarray}
{\cal L}_k [ n_+(.) ,n_-(.) , n ] && = k \lambda [ n_{+}(.); n_{-}(.) ] -   J^{geo}[ n_{+}(.); n_{-}(.) ] 
 \nonumber \\ &&
+ \chi_+ \left( n- \sum_{l} n_{+} \left( l \right)    \right)
+ \chi_- \left( n-  \sum_{l} n_{-} \left( l \right)  \right)
+\varphi \left( 1 - \sum_{l} l \left[ n_{+} \left( l \right)  + n_{-} \left( l \right) \right] \right)
\nonumber \\
&& =\sum_{l} n_+(l) \left[ k \ln \left\vert \theta_+(l) \right\vert - \ln \left( \frac{n_{+} (l) }{ p^{geo}_{+}(  l)  }    \right)
\right]
+ \sum_{l} n_-(l) \left[ k \ln \left\vert \theta_-(l) \right\vert - \ln \left( \frac{n_{-} (l) }{ p^{geo}_{-}(  l)  }   \right)
\right]
+ 2n \ln(n) 
 \nonumber \\ &&
 + \chi_+ \left( n- \sum_{l} n_{+} \left( l \right)    \right)
+ \chi_- \left( n-  \sum_{l} n_{-} \left( l \right)  \right)
+\varphi \left( 1 - \sum_{l} l \left[ n_{+} \left( l \right)  + n_{-} \left( l \right) \right] \right)
\label{foncq}
\end{eqnarray}

The optimization with respect to the empirical 1-interval observable $ n_{\pm}(l)  $ 
\begin{eqnarray}
0 = \frac{  \partial {\cal L}_k [ n_+(.) ,n_-(.) , n ] } {  \partial n_{\pm}(l)    } 
&& = -1 + k \ln \left\vert \theta_{\pm}(l) \right\vert - \ln \left( \frac{n_{\pm} (l) }{ p^{geo}_{\pm}(  l)  }    \right)
- \chi_{\pm} - \varphi l 
\label{deri2}
\end{eqnarray}
yields the forms
\begin{eqnarray}
 n_{\pm}(l)   = e^{ -1 - \chi_{\pm}  } 
 \left\vert \theta_{\pm}(l) \right\vert^k  p^{geo}_{\pm}(l)  e^{  - \varphi      l }
\label{pideri2}
\end{eqnarray}
The constraints
\begin{eqnarray}
n = \sum_{l} n_{\pm}(l) = e^{ -1 - \chi_{\pm}  } 
\sum_{l} \left\vert \theta_{\pm}(l) \right\vert^k  p^{geo}_{\pm}(l)  e^{  - \varphi      l }
\label{constraint1}
\end{eqnarray}
 determine the Lagrange multipliers $\chi_{\pm} $ as a function of the other parameters
\begin{eqnarray}
 e^{ -1 - \chi_{\pm}  }  = \frac{ n } {
\sum_{l} \left\vert \theta_{\pm}(l) \right\vert^k  p^{geo}_{\pm}(l)  e^{  - \varphi      l } }
\label{chi}
\end{eqnarray}

The optimization with respect to the interval density $n $
\begin{eqnarray}
0 = \frac{  \partial {\cal L}_k [ n_+(.) ,n_-(.) , n ]   } {  \partial n } && =
 2+  2 \ln(n)  + \chi_+  + \chi_- 
\label{deri1}
\end{eqnarray}
yields together with Eq. \ref{chi} that the value of the Lagrange multiplier $\varphi$ is fixed by the condition
\begin{eqnarray}
 1 = \left[ \sum_{l} \left\vert \theta_{+}(l) \right\vert^k  p^{geo}_{+}(l)  e^{  - \varphi      l }  \right]
 \left[ \sum_{l'} \left\vert \theta_{-}(l') \right\vert^k  p^{geo}_{-}(l')  e^{  - \varphi      l' }  \right]
\label{eqvarphi}
\end{eqnarray}
while the remaining constraint
\begin{eqnarray}
 1 = \sum_{l} l \left[ n_{+} \left( l \right)  + n_{-} \left( l \right) \right]  
\label{constraint2}
\end{eqnarray}
determines the value of the density $n$.

The value of the functional of Eq. \ref{foncq} for the optimal solution satisfying the constraints 
\begin{eqnarray}
{\cal L}_k^{opt} && =\sum_{l} n_+(l) \left[ k \ln \left\vert \theta_+(l) \right\vert - \ln \left( \frac{n_{+} (l) }{ p^{geo}_{+}(  l)  }    \right)
\right]
+ \sum_{l} n_-(l) \left[ k \ln \left\vert \theta_-(l) \right\vert - \ln \left( \frac{n_{-} (l) }{ p^{geo}_{-}(  l)  }   \right)
\right]
+ 2n \ln(n) 
\nonumber \\
&& =\sum_{l} n_+(l) \left[ 1 +\chi_+ + \varphi l   \right]
+ \sum_{l} n_-(l) \left[ 1 +\chi_- + \varphi l    \right]
+ 2n \ln(n) 
\nonumber \\
&& = n ( 2 + \chi_+ + \chi_- + 2  \ln(n) ) + \varphi 
\sum_{l} l \left[ n_+(l) + n_-(l)    \right] = \varphi 
\label{foncqopt}
\end{eqnarray}
actually reduces to the Lagrange multiplier $\varphi  $.

In summary, the scaled cumulant generating function $\varphi (k) $
 governing the exponential growth of the moments of Eq. \ref{thetaqav}
\begin{eqnarray}
\overline{ \left\vert \Theta_L  \right\vert^k } && = 
  \opsimeq_{L \to +\infty} 
 e^{   L \varphi (k)  }
\label{thetaqavgrowth}
\end{eqnarray}
 is the solution of Eq \ref{eqvarphi}
\begin{eqnarray}
 1 = \left[ \sum_{l=1}^{+\infty} \left\vert \theta_{+}(l) \right\vert^k  p^{geo}_{+}(l)  e^{  - l \varphi (k) }  \right]
 \left[ \sum_{l'=1}^{+\infty} \left\vert \theta_{-}(l') \right\vert^k  p^{geo}_{-}(l')  e^{  -  l' \varphi (k) }  \right]
\label{zeta}
\end{eqnarray}
that involves the distribution $p^{geo}_{\pm}(l) $ of the lengths of the intervals of the disorder configurations (Eq \ref{ppmgeom})
and the functions $\theta_{\pm}(l)$ of Eq. \ref{zprodinterval} of the observable under study.
One can check that the expansion at first order in $k$ around $k=0$ with Eq. \ref{typq0}
\begin{eqnarray}
\varphi (k) \opsimeq_{k \to 0} k \lambda^{typ}
\label{varphidv}
\end{eqnarray}
allows to recover the typical value $\lambda^{typ}$ de Eq \ref{lyatyp},
while the special case $\theta^{\pm}(l) = (t_{\pm})^l $ allows to recover the 
scaled cumulant generating function $\phi(k)  $ of Eq \ref{phikp} concerning the simpler case of products of random variables.


\begin{thebibliography}{99}


\bibitem{oono}
Y. Oono,
Progress of Theoretical Physics Supplement 99, 165 (1989).

\bibitem{ellis}
R.S. Ellis, Physica D 133, 106 (1999).

\bibitem{review_Touchette}
H. Touchette, Phys. Rep. 478, 1 (2009).

\bibitem{derrida-lecture}
B. Derrida, J. Stat. Mech. P07023 (2007).

\bibitem{harris_Schu}
R J Harris and G M Sch\"utz,
J. Stat. Mech.  P07020 (2007).

\bibitem{searles}
E.M. Sevick, R. Prabhakar, S. R. Williams, D. J. Searles,
Ann. Rev. of Phys. Chem.  Vol 59, 603 (2008). 

\bibitem{harris}
H. Touchette and R.J. Harris, chapter "Large deviation approach to nonequilibrium systems"
of the book "Nonequilibrium Statistical Physics of Small Systems: Fluctuation Relations and Beyond", Wiley 2013.

\bibitem{mft}
L. Bertini, A. De Sole, D. Gabrielli, G. Jona-Lasinio, and C. Landim
Rev. Mod. Phys. 87, 593 (2015).


\bibitem{lazarescu_companion}
A. Lazarescu, J. Phys. A: Math. Theor. 48 503001 (2015).

\bibitem{lazarescu_generic}
A. Lazarescu, J. Phys. A: Math. Theor. 50 254004 (2017).

\bibitem{fortelle_thesis}
A. de La Fortelle, PhD (2000).

\bibitem{vivien_Thesis}
V. Lecomte, PhD Thesis (2007).

\bibitem{chetrite_Thesis}
R. Ch\'etrite, PhD Thesis (2008).

\bibitem{wynants_Thesis}
B. Wynants, PhD Thesis (2010), arXiv:1011.4210.


\bibitem{chetrite_HDR}
R. Ch\'etrite, HDR Thesis (2018).





\bibitem{luckbook}
J.M. Luck, Al\'ea Saclay (1992)
"Syst\`emes d\'esordonn\'es unidimensionnels".


\bibitem{crisantibook}
A. Crisanti, G. Paladin, A. Vulpiani,
``Products of random matrices in statistical physics", Springer Verlag (1993). 



\bibitem{lifshitz64}
I.M. Lifshitz, Adv. Phys. 13 (1964) 483.

\bibitem{lifshitz65}
I.M. Lifshitz, Sov. Phys. Usp. 7 (1965) 549. 

\bibitem{lifbook}
I.M. Lifshitz, S.A. Gredeskul and L.A. Pastur, John Wiley and Sons (1987),
``Introduction to the theory of disordered systems".

\bibitem{boris}
B.I. Shklovskii and A.L. Efros, "Electronic properties of doped semiconductors",
Springer Verlag Belin (1984).


\bibitem{griffiths} 
R. B. Griffiths, Phys. Rev. Lett. 23, 17 (1969).

\bibitem{kafri} 
Y. Kafri and D. Mukamel, Phys. Rev. Lett. 91, 055502 (2003).

\bibitem{randeira}
M. Randeria, J. P. Sethna and R. G. Palmer 
Phys. Rev. Lett. 54, 1321 (1985) 

\bibitem{bray1}
A.J. Bray, Phys. Rev. Lett. 59, 586 (1987).

\bibitem{bray2}
A.J. Bray, Phys. Rev. Lett. 60, 720 (1988).

\bibitem{sdrgreview}
F. Igl\'oi and C. Monthus, Phys. Rep. 412, 277 (2005).

\bibitem{sdrgminireview}
F. Igl\'oi and C. Monthus, Eur. Phys. J. B 91, 290 (2018).

\bibitem{janssenrevue}
M. Janssen, Phys. Rep. 295, 1 (1998).

\bibitem{mirlinrevue}
F. Evers and A.D. Mirlin, Rev. Mod. Phys. 80, 1355 (2008).


\bibitem{Ludwig}
A.W.W. Ludwig, Nucl. Phys. B 330, 639 (1990).

\bibitem{Jac_Car}
J.L. Jacobsen and J.L. Cardy, Nucl. Phys., B515, 701 (1998).

\bibitem{Cha_Ber}
C. Chatelain and B. Berche, Nucl. Phys., B572, 626 (2000).

\bibitem{Cha_Ber_Sh}
C. Chatelain, B. Berche and L.N. Shchur, J. Phys. A Math. Gen. 34, 9593 (2001).

\bibitem{PCBI}
G. Pal\'agyi, C. Chatelain, B, Berche and F. Igl\'oi, Eur. Phys. J.
B13, 357 (2000).

\bibitem{BCrevue}
B. Berche and C. Chatelain, in {\it Order, disorder, and criticality}, ed. 
by Yu. Holovatch, World Scientific, Singapore 2004, p.146. 



\bibitem{Thi_Hil}
M.J. Thill and H.J. Hilhorst, J. Phys. I 6, 67 (1996).

\bibitem{symetriemultif}
 C. Monthus, B. Berche and C. Chatelain, 
J Stat. Mech.  P12002 (2009)



\bibitem{ellisbook}
R.S. Ellis, "Entropy, Large Deviations, and Statistical Mechanics",
 Springer Verlag New-York (1985).


\bibitem{deuschel}
J.D. Deuschel and D.W. Stroock, "Large Deviations", Academic Press Boston (1989).

\bibitem{dembo}
A. Dembo and O. Zeitouni,
"Large Deviations Techniques and Applications", Springer Verlag Berlin (1998).


\bibitem{denHollander}
F. den Hollander, "Large Deviations", Fields Institute Monographs,
American Mathematical Society, Providence (2008).

\bibitem{saintflour}
R. Azencott, M.I. Freidlin and S.R.S. Varadhan,
"Large Deviations at Saint-Flour ", Probability at Saint-Flour, Springer Heidelberg (2012)

\bibitem{timo}
F. Rassoul-Agha and T. Seppalainen,
"A Course on Large Deviations with an Introduction to Gibbs Measures", Graduate Studies in Mathematics volume 162,
American Mathematical Society (2015)



\bibitem{talagrand}
M. Talagrand, "Spin-glasses : a challenge for mathematicians", springer New-York (2003)

\bibitem{bovier}
A. Bovier,  "Statistical mechanics of disordered systems : a mathematical perspective",
Cambridge University Press, Cambridge (2006).

\bibitem{comets}
F. Comets,
"Directed Polymers in Random Environments", Probability in Saint-Flour, 
Lecture Notes in Mathematics 2175, Springer International Publishing  (2017).

\bibitem{zeitouni}
O. Zeitouni, J. Phys. A Math. Gen. 39, R433 (2006).







\bibitem{tourigny}
A. Comtet and Y. Tourigny, arxiv: 1601.01822




\bibitem{Der_Hil_corre}
B. Derrida and H. Hilhorst, J. Phys. C Solid State Phys. 14, L539 (1981).

\bibitem{c_maj}
C. Monthus, J. Phys. A: Math. Theor. 51, 465301 (2018).


\bibitem{alt_levitov}
B.L. Altshuler, Y. Gefen, A. Kamenev and L.S. Levitov,
Phys. Rev. Lett. 78, 2803 (1997).

\bibitem{luca}
A. De Luca, B.L. Altshuler, V.E. Kravtsov and A. Scardicchio, Phys. Rev. Lett. 113, 046806 (2014).

\bibitem{forward}
F. Pietracaprina, V. Ros, A. Scardicchio, Phys. Rev. B 93, 054201 (2016).

\bibitem{c_mblfss}
C. Monthus, J. Stat. Mech. (2016) 123303

\bibitem{cavity1}
L.B. Ioffe and M. M\'ezard, Phys. Rev. Lett. 105, 037001 (2010).

\bibitem{cavity2}
M.V. Feigelman, L.B. Ioffe and M. M\'ezard, Phys. Rev. B 82, 184534 (2010).

\bibitem{cavity3}
O. Dimitrova and M. M\'ezard, J. Stat. Mech. P01020 (2011).







\bibitem{fortelle_chain}
G. Fayolle and A. de La Fortelle,
Problems of Information Transmission 38, 354 (2002).

\bibitem{fortelle_jump}
A. de La Fortelle, 
Problems of Information Transmission 37 , 120 (2001).



\bibitem{maes_canonical}
C. Maes and K. Netocny, Europhys. Lett. 82, 30003 (2008)

\bibitem{maes_onandbeyond}
C. Maes, K. Netocny and B. Wynants, Markov Proc. Rel. Fields. 14, 445 (2008).

\bibitem{wynants_thesis}
B. Wynants, PhD Thesis (2010), arXiv:1011.4210.

\bibitem{chetrite_formal}
A. C. Barato and R. Chetrite, J. Stat. Phys. 160, 1154 (2015).

\bibitem{BFG1}
L. Bertini, A. Faggionato and D. Gabrielli, 
Ann. Inst. Henri Poincare Prob. and Stat. 51, 867 (2015).

\bibitem{BFG2}
L. Bertini, A. Faggionato and D. Gabrielli, 
Stoch. Process. Appli. 125, 2786 (2015).


\bibitem{LargeDevRings}
 C. Monthus,  J. Stat. Mech. 023206 (2019).

\bibitem{largeDevOpen}
 C. Monthus, J. Phys. A: Math. Theor. 52, 025001 (2019).


\bibitem{mFT}
 C. Monthus, J. Phys. A: Math. Theor. 52 135003(2019).
 



\bibitem{maes_diffusion}
C. Maes, K. Netocny and B.  Wynants
Physica A 387, 2675 (2008).


\bibitem{engel}
J. Hoppenau, D. Nickelsen and A. Engel,
 New J. Phys. 18 083010 (2016).



\bibitem{Der_Spo}
B. Derrida and H. Spohn, J. Stat. Phys., {\bf 51}, 817 (1988).

\bibitem{Coo_Der}
J. Cook and B. Derrida, J. Stat. Phys. {\bf 63}, 505 (1991).

\bibitem{abou1}
R. Abou-Chacra, P.W. Anderson and D.J. Thouless, 
J. Phys. C : Solid State Physics 6, 1734 (1973).

\bibitem{abou2}
R. Abou-Chacra and  D. J. Thouless,
  J. Phys. C: Solid State Phys. 7, 65 (1974).


\bibitem{DR}
B. Derrida and G.J. Rodgers, J. Phys. A : Math. Gen. 26, L457 (1993).

\bibitem{MD}
J.D. Miller and B. Derrida, J. Stat. Phys. 75, 357 (1994).

\bibitem{rem}
B. Derrida, Phys. Rev. B 24, 2613 (1981).




\bibitem{Der_Tou}
B. Derrida and G. Toulouse, J. Phys. Lett. (France), {\bf 46}, L223
(1985).

\bibitem{Der_Fly}
B. Derrida and H. Flyvbjerg, J. Phys. A Math. Gen. 20, 5273 (1987).


\bibitem{Der_review}
B. Derrida, ``Non-self-averaging effects in sums of random variables, spin
glasses, random maps and random walks'', in ``On three levels''
 Eds M. Fannes et al (1994) New-York Plenum Press. 

\bibitem{gumbel}
E.J. Gumbel, `` Statistics of extreme'' (Columbia University Press, NY 1958).

\bibitem{galambos}
J. Galambos, `` The asymptotic theory of extreme order statistics'' 
( Krieger , Malabar, FL 1987).

\bibitem{levitov1}
L.S. Levitov, Europhys. Lett. 9, 83 (1989).

\bibitem{levitov2}
L.S. Levitov, Phys. Rev. Lett. 64, 547 (1990).

\bibitem{levitov3}
B.L. Altshuler and L.S. Levitov, Phys. Rep. 288, 487 (1997).

\bibitem{levitov4}
L.S. Levitov, Ann. Phys. (Leipzig) 8, 5, 507 (1999).

\bibitem{mirlin_evers}
 F. Evers and A. D. Mirlin
Phys. Rev. Lett. 84, 3690 (2000); \\
A.D. Mirlin and F. Evers, 
Phys. Rev. B 62,  7920 (2000).

\bibitem{fyodorov}
Y.V. Fyodorov, A. Ossipov and A. Rodriguez, J. Stat. Mech. L12001 (2009).

\bibitem{fyodorovrigorous}
Y.V. Fyodorov, A. Kupiainen and C. Webb. arxiv:1509.01366.


\bibitem{oleg1}
O. Yevtushenko and V. E. Kratsov,
J. Phys. A 36, 8265 (2003).

\bibitem{oleg2}
O. Yevtushenko and A. Ossipov, 
J. Phys. A 40, 4691 (2007).

\bibitem{oleg3}
S. Kronm\"uller, O. M.  Yevtushenko and E. Cuevas, 
J. Phys. A 43, 075001 (2010).

\bibitem{oleg4}
V. E. Kratsov, A. Ossipov, O. M.  Yevtushenko and E. Cuevas, 
Phys. Rev. B 82, 161102(R) (2010).

\bibitem{olivier_per}
E. Bogomolny and O. Giraud, Phys. Rev. E 84, 036212 (2012).

\bibitem{olivier_strong}
E. Bogomolny and O. Giraud, Phys. Rev. E 84, 046208 (2012).

\bibitem{olivier_conjecture}
E. Bogomolny and O. Giraud, Phys. Rev. Lett. 106, 044101 (2011).


\bibitem{us_strongmultif}
C. Monthus and T. Garel, J. Stat. Mech. (2010) P09015.

\bibitem{c_mblperturb}
C. Monthus, Entropy 18, 122 (2016).

\bibitem{c_mblstrongmultif}
C. Monthus, J. Stat. Mech.  073301 (2016).




\bibitem{mblcayley}
 C. Monthus,  J. Stat. Mech. 123304 (2017).



\bibitem{derrida_leb}
B. Derrida and J.L. Lebowitz, Phys. Rev. Lett. 80, 209 (1998)

\bibitem{dean_maj}
D.S. Dean and S.N. Majumdar,
Phys. Rev. Lett. 97, 160201 (2006) and
Phys. Rev. E 77, 041108 (2008).


\bibitem{maj_verg}
S. N. Majumdar and M. Vergassola,
Phys. Rev. Lett. 102, 060601 (2009).


\bibitem{largedevAsym}
C. Monthus, arxiv:1904.02448




\bibitem{gaspard}
D. Andrieux and P. Gaspard, JSTAT P11007 (2008).

\bibitem{maes_semi}
C. Maes, K.l Netocny and B. Wynants,
J. Phys. A: Math. Theor. 42 (2009) 365002


\bibitem{zambotti}
M. Mariani and L. Zambotti, Adv. Appl. Prob. 48, 648 (2016).

\bibitem{faggionato}
A. Faggionato, arxiv:1709.05653



\end{thebibliography}
\end{document}